\documentclass[3p]{elsarticle}
\usepackage{amsfonts}
\usepackage{amsmath}
\usepackage{lineno,hyperref}
\usepackage{graphicx}
\usepackage{subfigure}

\newtheorem{theorem}{Theorem}

\modulolinenumbers[5]

\journal{Biochemical Engineering Journal}
\begin{document}

\begin{frontmatter}

\title{Monitoring management criterion of an anaerobic digester using an explicit model based on temperature and pH}

\author[ama]{Ang\'elica M. Alzate Iba\~nez}
\address[ama]{Facultad de Ingenier\'ia y Arquitectura, Departamento de Ingenier\'ia Electrica, Electr\'onica y Computaci\'on,
Universidad Nacional de Colombia, Sede Manizales, Bloque Q, Campus La Nubia, Manizales 170003, Colombia.}

\author[com]{Carlos Ocampo-Martinez}
\address[com]{Universitat Polit\`{e}nica de Catalunya, Institut de Rob\`{o}tica i Inform\`{a}tica Industrial (CSIC-UPC), Llorens i Artigas 4-6, 08028, Barcelona, Spain}

\author[cac]{Carlos A. Cardona Alzate}
\address[cac]{Instituto de Biotecnolog\'ia y Agroindustria, Departamento de Ingenier\'ia Qu\'imica, Universidad Nacional de Colombia Sede Manizales, Cra. 27 No. 64-60, Manizales 170003, Colombia.}

\author[vmt]{V\'ictor M. Trejos M.}
\ead{victor.trejos@iquimica.unam.mx}
\address[vmt]{Instituto de Qu\'imica, Universidad Nacional Aut\'onoma de M\'exico,  
Circuito Exterior, 04510 M\'exico D.F., M\'exico.}
\cortext[jmt]{Corresponding author}


\begin{abstract}
In this paper, the stability analysis of an anaerobic digestion process is presented. The analysis is performed using a simplified mathematical model, which includes explicit temperature and pH dependence on kinetic growth rates. A detailed nonlinear and bifurcation analyses are performed in order to study the effects of temperature and pH parameters on the process behaviour. In addition, both safety and optimal operation regions of the bioreactor are established. Based on bifurcation diagrams, it is observed that the washout condition occurs by combining a fold bifurcation and a transcritical bifurcation. Consequently, to prevent the washout and guarantee optimal operation conditions of the bioreactor, a risk criterion oriented to monitor the bioprocess on-line is proposed, allowing to detect the system destabilisation.
\end{abstract}
\begin{keyword}
Anaerobic digestion, bioreactor, dynamic analysis, bifurcation theory, nonlinear mathematical model.
\end{keyword}

\end{frontmatter}


%
%
%
%
%

\subsection*{Nomenclature}
\noindent\begin{tabular}{lp{0.7\columnwidth}}
D   & dilution rate, day$^{-1}$\\
I$_{pH}$ & inhibition factor by pH, dimensionless \\
k$_1$ & yield for substrate degradation, g COD$/$g X$_1$ \\
k$_2$ & yield for VFA production, mmol VFA$/$g X$_1$\\
k$_3$ & yield for VFA consumption, mmol VFA$/$g X$_2$\\
K$_{1}$ & half-saturation constant associated with $S_1$, g$/$L\\
K$_{2}$ & half-saturation constant associated with $S_2$, mmol$/$L\\
K$_{I}$ & inhibition constant associated with $S_2$, mmol$^{2}$ $/$L $^{2}$ \\
pH & hydrogen potential, dimensionless \\
pH$_{UL}$ & upper limit of pH, dimensionless \\
pH$_{LL}$ & lower limit of pH, dimensionless \\
$T$ & operational temperature, $^ \circ$ C\\
$T_{\text{min}}$ & minimum temperature, $^ \circ$ C \\
$T_{\text{max}}$ & maximum temperature, $^ \circ$ C \\ 
$T_{opt}$ & optimal temperature, $^ \circ$ C \\
S$_1$ & outlet organic substrate concentration, g L$^{-1}$\\
S$_1^0$ & inlet organic substrate concentration, g L$^{-1}$\\
S$_2$ & outlet volatile fatty acid concentration, mmol L$^{-1}$\\
S$_2^0$ & inlet volatile fatty acid concentration, mmol L$^{-1}$\\
X$_1^0$ & inlet acidogenic biomass concentration, g L$^{-1}$\\
X$_1$ & acidogenic biomass concentration, g L$^{-1}$\\
X$_2^0$ & inlet methanogenic biomass concentration, g L$^{-1}$\\
X$_2$ & methanogenic biomass concentration, g L$^{-1}$\\
$\alpha$ & proportion of dilution rate for bacteria, dimensionless\\
$\bar \mu_1$ & acidogenic growth rate kinetic modified, day$^{-1}$ \\ 
$\bar \mu_2$ & methanogenic growth rate kinetic modified, day$^{-1}$ \\ 
${\mu _{1\max }}$ & maximum value for acidogenic growth rate kinetic, day$^{-1}$ \\ 
${\mu _{2\max }}$ & maximum value for methanogenic growth rate kinetic, day$^{-1}$ \\ 
$\Theta$ & temperature activity coefficient, dimensionless
\end{tabular}

\section{INTRODUCTION}
The growing interest for the protection of the environment leads to the development of more detailed studies dealing with the analysis of systems used in the wastewater treatment. Anaerobic digestion is one of the most common biological process in the industrial applications and municipal wastewater treatment \cite{Sbarciog10}. Since the eighties, industrial wastewater treated by anaerobic digestion began to grow and extend worldwide, such that, in 2007, the overall number of anaerobic reactors treating industrial wastewater reached more that two thousand references and kept on increasing since then \cite{Jimenez15}.

The anaerobic digestion process is an attractive waste treatment that involves degradation of organic materials by the action of microbial populations. 
This process is characterized by the existence of highly non-linear dynamics and its efficient operation is affected by several factors such as: load disturbances, system uncertainties, limited online measurement information, constraints on variables, and uncertain kinetics \cite{Mendez10}. The last factor is strongly affected by temperature and pH, both being key for biomass growth during the anaerobic digestion process. Hence, the understanding of the complex nonlinear behaviour of the anaerobic digestion system allows to give a qualitative description of the stability of the system.

In the last decade, dynamic analysis of anaerobic digestion process has been studied. Shen \textit{et al.} \cite{Shen07} have investigated the stability of stationary solutions of an anaerobic digestion model. The authors argued that the dynamics of the system depend on kinetic model and key parameters such as dilution rate.
 
Also, the authors provided a guidance for anaerobic digestion reactor design, operation and control in terms of robustness and productivity. However, the effects of biomass concentrations and products on the dynamics were not reported. In 2008, Hess and Bernard \cite{Hess08} proposed a simplified criterion to evaluate the operational risk from the analysis of the nonlinear system. This criterion showed a suitable performance in a real plant. When applied, it was able to diagnose an operation strategy by detecting an early destabilization caused by accumulation of volatile fatty acids. Later, Hess \cite{Hess09} presented an extension of the work in \cite{Hess08}. The authors demonstrated with real data the efficiency of the criterion and also proposed a methodology to monitor in real-time the trajectory of the system.

The results were obtained from a general approach, thus the methodology could be applied even without knowing the model parameters. Dimitrova and Krastanov \cite{Dimitrova09} studied the equilibrium points stability and performed a bifurcation analysis of the model varying the dilution rate parameter in order to study the system in open loop, and proposed a feedback control law for asymptotic stabilization of the closed-loop system. Rinc{\'o}n \textit{et al.} \citep{Rincon09} studied an open-loop dynamic analysis and stability of the system varying the dilution rate parameter. The results show that the washout condition occurs after a combination of both fold and transcritical bifurcations. They also used their obtained results to develop an adaptive control strategy. Sbarciog \textit{et al.} \cite{Sbarciog10} proposed a methodology to estimate the separatrix between the stable attraction basins
of the equilibria. Their analysis was based on growth kinetic models with and without inhibition and the results obtained from the inputs and initial states led to achieve the proper operation of the system. Recently, Benyahia \textit{et al.} \cite{Benyahia12} proposed a generic methodology for the stability analysis of a two-step bioprocess system. 

Dynamic analysis of anaerobic systems based on simplified mathematical models have been an important and active research area to determine better operational conditions, on-line monitoring, and design control strategies \cite{Benyahia12}. The simplified mathematical models assume a rate-limiting step, and these have been widely used due to their \textquotedblleft  relative simplicity and high capacity in order to reproduce the dynamical behaviour of main operational parameters of the process\textquotedblright \cite{Benyahia12} and computational easiness. Nowadays, simplified models have been positioned as a suitable alternative to describe biotechnological processes \cite{Donoso11}. Initially, simplified models of anaerobic systems have been studied using specific kinetics with limited number of functioning conditions. 
Recently, research of anaerobic systems is focused on the analysis of the equilibria and local stability of the system against variations in the dilution factor. Nevertheless, the anaerobic digestion process is affected by other factors, e.g., ammonia concentration, acclimation, sulfate, heavy metals, pH, temperature, presence of other ions \cite{Chen08}. However, the dynamical analysis of anaerobic digestion systems including factors such as inhibitory effects of temperature and pH on the kinetics rates have not been considered yet. 

Based on the background described above and the knowledge gap identified, the purpose of the this work is to determine optimal operating conditions by analysing an anaerobic digestion process under variations in dilution rate, temperature and pH. In this way, equilibria and stability analysis using bifurcation theory are performed at different operational conditions. A simplified mathematical model proposed by Bernard \textit{et al.} \citep{Bernard01} was considered, assuming  that the process occurs in two main stages: acidogenesis and methanogenesis. Also, mathematical expressions of temperature and pH activity coefficients are included in the Monod and Haldane 
bacterial growth functions, the CTM model \citep{Rosso93} for temperature and Angelidaki expression \citep{Angelidaki93} for pH inhibition. The influence of temperature and pH on the kinetic reactions and the stability of the equilibrium points were determined. To take this analysis into account, an operational indicator of the processes is proposed, which could be used for on-line monitoring from the measurement of dilution factor, temperature, and pH parameters. These criterion allows to identify in a preventive way a possible destabilization of the bioreactor.

The remainder of this paper is organized as follows. Sec. \ref{SectionII} describes the anaerobic digestion process assuming two main stages, acidogenic and methanogenic. The mathematical model accounting for the effects of temperature and pH  inhibition on growth rates is discussed. Sec. \ref{SectionIII} details the bifurcation parameters, operational methodology and the software packages used. Then, results are presented and discussed in Sec. \ref{SectionIV}. The analytical equilibrium points are presented in Sec. \ref{SectionIVa}. The dynamic analysis using bifurcation theory, and the qualitative analysis of the system are discussed in Sec. \ref{SectionIVb}. Sec. \ref{SectionIVc} presents the results and the global behaviour of the system oriented to on-line monitoring, and a risk monitoring management is proposed and evaluated in Sec. \ref{SectionIVd}. Finally, Sec. \ref{SectionV} provides the main conclusions of this work. 
\section{MATHEMATICAL MODEL} \label{SectionII}
\begin{figure*}[h]
  \centering
  \includegraphics[width=13cm,height=10cm]{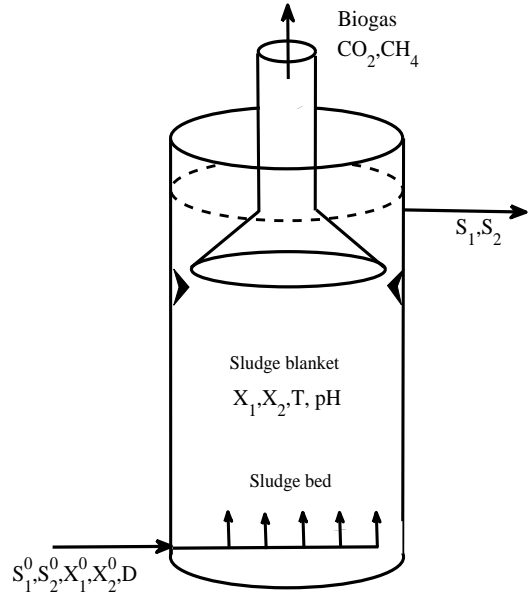}
  \caption{Schematic diagram of the considered UASB reactor.}
  \label{fig1}
\end{figure*}

The mathematical model proposed by Bernard \textit{et al.} \cite{Bernard01} is considered in this paper. The model assumes that the anaerobic digestion process is based on two main reactions: acidogenic and methanogenic. Acidogenesis is the stage in which organic substrates $S_1$ are catabolized by acidogenic bacteria $X_1$, generating intermediates compounds with lower molecular weight, such as Volatile Fatty Acids (VFA) concentration $S_2$. Then, $S_2$ is degraded by methanogenic bacteria $X_2$ into methane and carbon dioxide. The anaerobic digestion scheme of the process is shown in Fig. \ref{fig1}. The mathematical model by component in the liquid phase is given by
\begin{subequations} \label{eq1}
\begin{flalign}
\dot{X_1} & = D\left( {X_1^0 - \alpha {X_1}} \right) + {\mu _1}{X_1}, \label{eq3}\\
\dot{S_1} & = D\left( {S_1^0 - {S_1}} \right) - {k_1}{\mu _1}{X_1}, \label{eq4}\\
\dot{X_2} & = D\left( {X_2^0 - \alpha {X_2}} \right) + {\mu _2}{X_2}, \label{eq5}\\
\dot{S_2} & = D\left( {S_2^0 - {S_2}} \right) + {k_2}{\mu _1}{X_1} - {k_3}{\mu _2}{X_2}, \label{eq6}
\end{flalign}
\end{subequations}
where $D$ is the dilution rate, $\alpha$ is the fraction of bacterias in the liquid phase, $k_1$ is the yield for substrate degradation, $k_2$ is the yield for VFA production, $k_3$ is the yield for VFA consumption, and the superscript ($^0$) refers to the initial input concentration. Besides, $\mu_{1}$ and $\mu_{2}$ are the 
Monod and Haldane growth rate kinetic model, respectively \cite{Monod50,Haldane01}. 

Both temperature and pH parameters affect directly the survival of bacterial population. Low values of pH are an usual indicator of the process destabilization by VFA accumulation. Also, a gas solubility and microbial growth rates are directly affected by the temperature. A decreased temperature process reduces the metabolic
activity of methanogenic bacteria. On the other hand, an increased temperature process takes the advantage of stimulating the rate of the microorganisms but also leads higher concentration of ammonia and gases. In this paper, the growth rate kinetic models were modified as a function of temperature and pH as follows.
\subsection{Growth rate kinetics}
Mathematical expressions of kinetic models including the effect of temperature and pH are given by
\begin{align}
{\bar \mu _1} & = \mu_{1\max}\left(  \frac{{{S_1}}}{{{K_1} + {S_1}}} \right) 
\Theta {I_{pH}}, \label{Monod1}\\
{\bar \mu _2} & = {\mu _{2\max }}
\left(
\frac{S_2}{K_2+S_2+\frac{S_2^2}{K_I}}
\right)
\Theta {I_{pH}}, \label{Haldane1}
\end{align}
where $\bar \mu_1$ is Monod modified, $\bar \mu_2$ Haldane modified, $\Theta$ is the CTM temperature coefficient, $I_{pH}$ is the inhibition factor by pH, $K_{1}$ is the half-saturation constant associated with $S_1$, $K_{2}$ is the half-saturation constant associated with $S_2$, and K$_{I}$ is the inhibition constant associated
with $S_2$. Finally, $\mu_{1\max}$ and $\mu_{2\max}$ denotes the maximum value for the acidogenic and methanogenic growth rates, respectively.
\subsubsection{Temperature activity coefficient ($\Theta$)}
Several authors have found that the effect of temperature on kinetic parameters can be easily modelled by means of theArrhenius's equation \cite{Crites98,Banik98,Yuan11}. However, some disadvantages are observed at each step of the microbial population \cite{Pavlostathis91}. In this way, Rosso {\it et al.} \cite{Rosso93} proposed a suitable temperature model in order to describe the temperature influence in an anaerobic process. This model has been referred in the literature as a cardinal temperature model (CTM). The applicability of CTM in anaerobic processes has been shown by Donoso-Bravo {\it et al.} \cite{Donoso13}. The CTM temperature activity 
coefficient $\Theta$ is given by
\begin{equation} \label{ITemperature}
   \Theta={\frac{(T-T_{max})(T-T_{\text{min}})^2}
  {(T_{opt}-T_{min})
  (T_b-T_a)}},
\end{equation}
where $T_a=(T_{opt}-T_{max})(T_{opt} +T_{min}-2T)$, $T_b=(T_{opt}-T_{min})(T-T_{opt})$, ${T}$ is the 
operation temperature, $T_{\text{min}}$ and $T_{\text{max}}$ are the lower and upper temperatures where the growth rate does not occurs, respectively, and $T_{opt}$ is the temperature at which the maximum specific growth rate equals its optimal value. The optimal temperature range for the bacterial microorganism activity has been established in the range $[25,35]^{\circ}$C according to Tchobanoglous \cite{Tchobanoglous03}. The metabolic activity of 
methanogenic bacteria is reduced significantly at the psychrophilic temperature range. In some cases, the growth rate activity can be inactivated ($T<15^{\circ}$C). 
\subsubsection{pH inhibition factor ($I_{pH}$)}
In anaerobic digestion processes, the pH affects the growth of microorganisms as well as the bioreactor performance, since the presence of ammonia concentration, acclimation, sulphate, and VFA leads to acidity or alkalinity and determines the kind of microorganisms in the anaerobic digester. The pH inhibition factor $I_{pH}$ is given by \cite{Batstone02,Angelidaki93}
\begin{equation} \label{IpH}
   {I_{pH}} = {\frac{{1 + 2 \times {{10}^{0.5(p{H_{LL}} - p{H_{UL}})}}}}{{1 + 
   {{10}^{(pH - p{H_{UL}})}} + {{10}^{(p{H_{LL}} - pH)}}}}},
\end{equation}
where, $p{H_{UL}}$ and $p{H_{LL}}$ are the upper and lower limits of pH, when the specific microbial growth rate is reduced 50\% of its initial value without inhibition, respectively. The optimal pH range for the microorganisms activity according to Sanchez \cite{Sanchez00} is given between 6.8 and 7.4. At this pH range, the tolerance of anaerobic microorganisms in anaerobic process is better, while it favours optimum operation condition of methanogenic bacteria growth. At lower pH values ({\it i.e.}, pH $\leq$ 4), the VFA production and methanogenic bacteria inhibition occurs by acidifying and, at higher pH values ({\it i.e.}, pH $\geq$ 8.2), the activity inhibition of the bacterial populations is presented in the anaerobic process.
\section{MATERIALS AND METHODS}\label{SectionIII}
Although the behaviour of the Monod and Haldane kinetic models is well known in 
differents aplications as fermentative processes \cite{Trejos09}, in this paper the behaviour of these models under the effects of temperature and pH was analysed and discussed.  In addition, several authors \cite{Hess08,Rincon09,Benyahia12} have reported the analysis of the existence and stability of the equilibrium points of Bernard's model \citep{Bernard01}. However, the dynamic analysis of the system in function of the temperature and pH has not yet been reported. Here, analytical equilibrium points of the system are computed taking into account the effects of 
temperature and pH on the growth rates expressions. The equilibria analysis of the four-dimensional state-space system is given by doing the left-hand side of \eqref{eq1} equal to zero  ($\dot{X_1}=\dot{X_2}=\dot{S_1}=\dot{S_2}=0$). Then, a dynamical analysis of the system and a sensitive numerical analysis 
as a function of $D$, $T$ and pH parameters are performed.

The bifurcation analysis of the system is carried out using bifurcation theory. The software package MATCONT \cite{Dhooge06} is used for numerical simulations.MATCONT solves the equilibrium values of $X_1$, $S_1$, $X_2$ and $S_2$ from differential equations in \eqref{eq1} by continuation from an starting point satisfying the condition $\dot{X_1}=\dot{S_1}=\dot{X_2}=\dot{S_2}=0$. First, all the parameters were keep fixed other than $D$, which was taken as a bifurcation parameter. Subsequently, the same procedure was performed taking $T$ and pH as a bifurcation parameters.

The bifurcation diagrams were analysed by regions in order to study the number and nature of the equilibria in the range of the operating parameters. The stability analysis around the equilibrium point was carried out by the Lyapunov's indirect method \cite{Slotine91}. The result of its application leads to a necessary but not a sufficient conditions for the stability of the system. Therefore, the Hurwitz criterion \cite{Allen07} was applied in order to ensure it (see \ref{Stability}).

Once the system has been described using the bifurcation theory with the considered bifurcation parameters ($D$, $T$ and pH), the study of was focused on the identification of critical values and operational optimal conditions. The global
behaviour of the system is studied by stages involving the effects of temperature and pH using the operational ranges defined by the bifurcation analysis. As mentioned by some authors, the acidogenic stage can be studied independently of the methanogenic stage \cite{Rincon09, Benyahia12, Hess08}. Then, the behaviour of each stage was studied separately based on dynamical analysis results, and  the conditions that ensure the stability of the system were established.

Finally, the last step in the proposed analysis was the establishment of a criterion to monitor the bioreactor online. A risk index was proposed in order to monitor online the process. Numerical simulations using the experimental data reported by Bernard \textit {et al.} \cite{Bernard01} in an anaerobic upflow were performed, with the purpose of evaluating the risk-index applicability. All the results are presented in Section \ref{SectionIV}.
\section{RESULTS}\label{SectionIV}
\subsection{Effects of pH and T on growth rate kinetics}
\begin{figure*}[h]
  \centering
  \subfigure{\label{Fig2a}
  \hspace*{-1.7cm}\includegraphics[width=7.8cm,height=7cm]{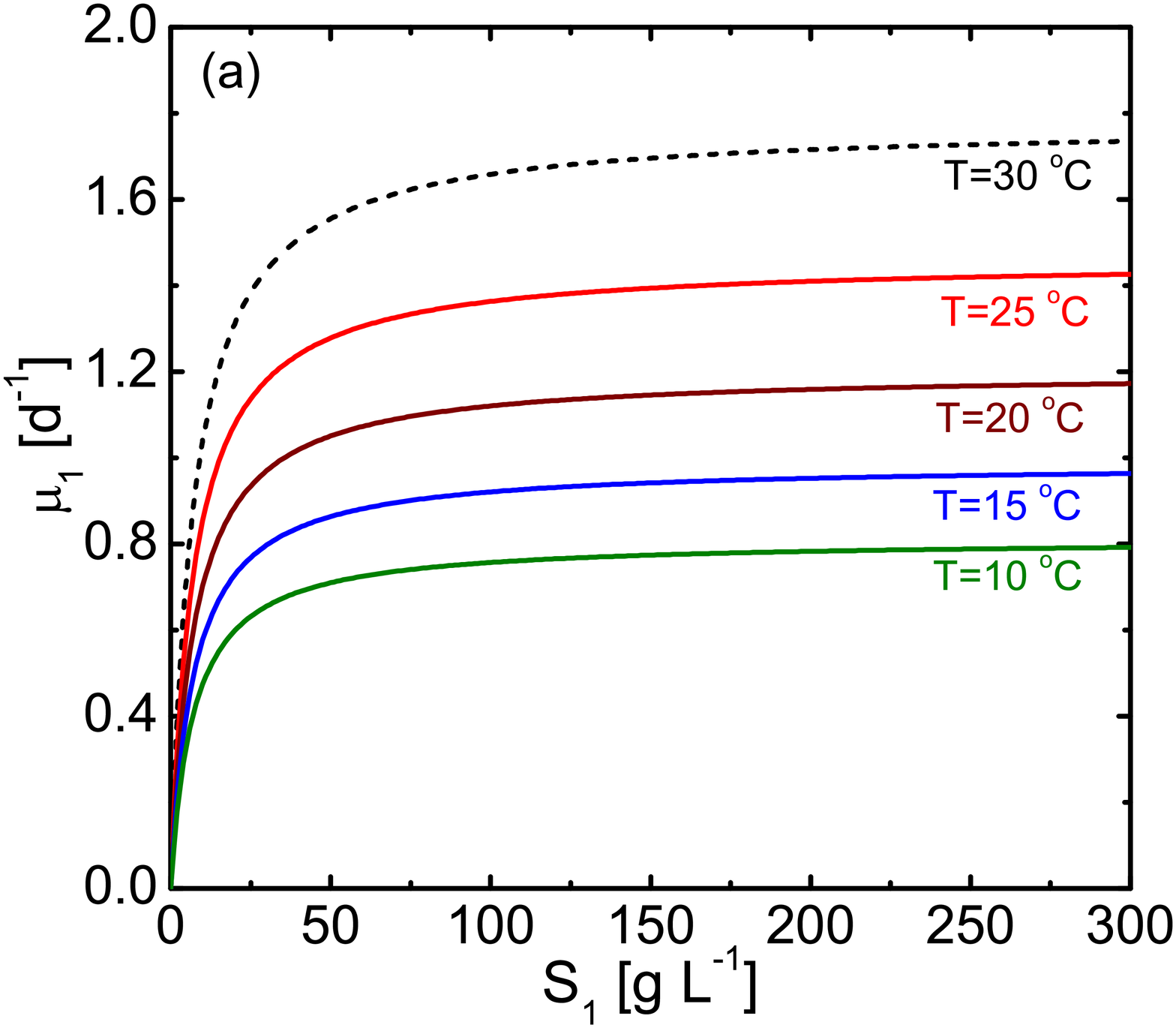}} 
  \subfigure{\label{Fig2b}
  \hspace*{-0.3cm}\includegraphics[width=7.8cm,height=7cm]{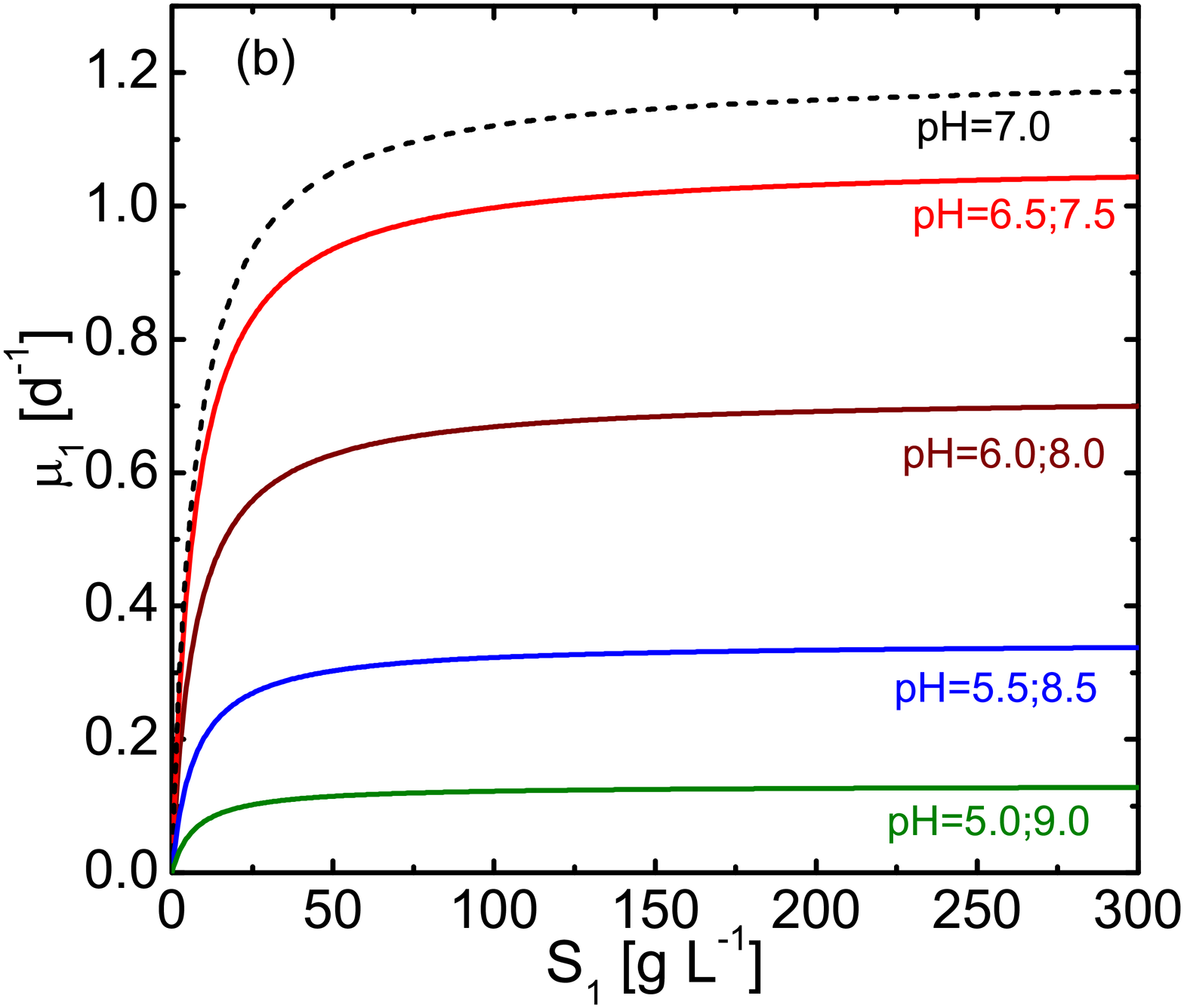}}
  \caption{
  Monod growth rate ($\mu_1$) as a function of the organic substrate $S_1$.
  Cases presented:
  (a) changing with the temperature factor $\Theta$,  and 
  (b) changing with the pH factor $I_{\text{pH}}$.  
  The dotted line shows the behaviour of the maximum kinetic growth rates 
  at optimal temperature and pH ({\it i.e.}, at T=30$^\circ$C and pH=7). 
  }
  \label{Fig2}
\end{figure*}

For the sake of simplicity, identical notation in all figures is used. In Fig. \ref{Fig2}, the influence of temperature and pH in the microbial Monod growth rate ($\bar \mu_1$) as a function of the organic substrate $S_1$ is shown. As can be observed in this figure, the growth rate is an increasing function when ${S_1} \ge 0$, and it exhibits a maximum value at ${\bar \mu _1}\left( {+\infty } \right) = {\mu _{1\max}}$. In Fig. \ref{Fig2a}, the behaviour of the growth rate model is obtained in a range of temperature between 10$^\circ$C and 30$^\circ$C. As can be observed in this figure, at low temperatures the growth rate is reduced dramatically and at high temperatures as 30$^\circ$C, the maximum curve of the growth rate is obtained. The growth rate as a function of the pH is shown in Fig. \ref{Fig2b}. The numerical simulations are performed in the range of pH 5.0-9.0 in order to compare the bioreactor performance at different pH values. In the case of pH, as values move away from neutral pH (pH=7), a remarkable decrease of the growth rate is observed through the organic substrate values $S_1$.

\begin{figure*}[h]
  \centering
  \subfigure{\label{Fig3a}
  \hspace*{-1.7cm}\includegraphics[width=7.8cm,height=7cm]{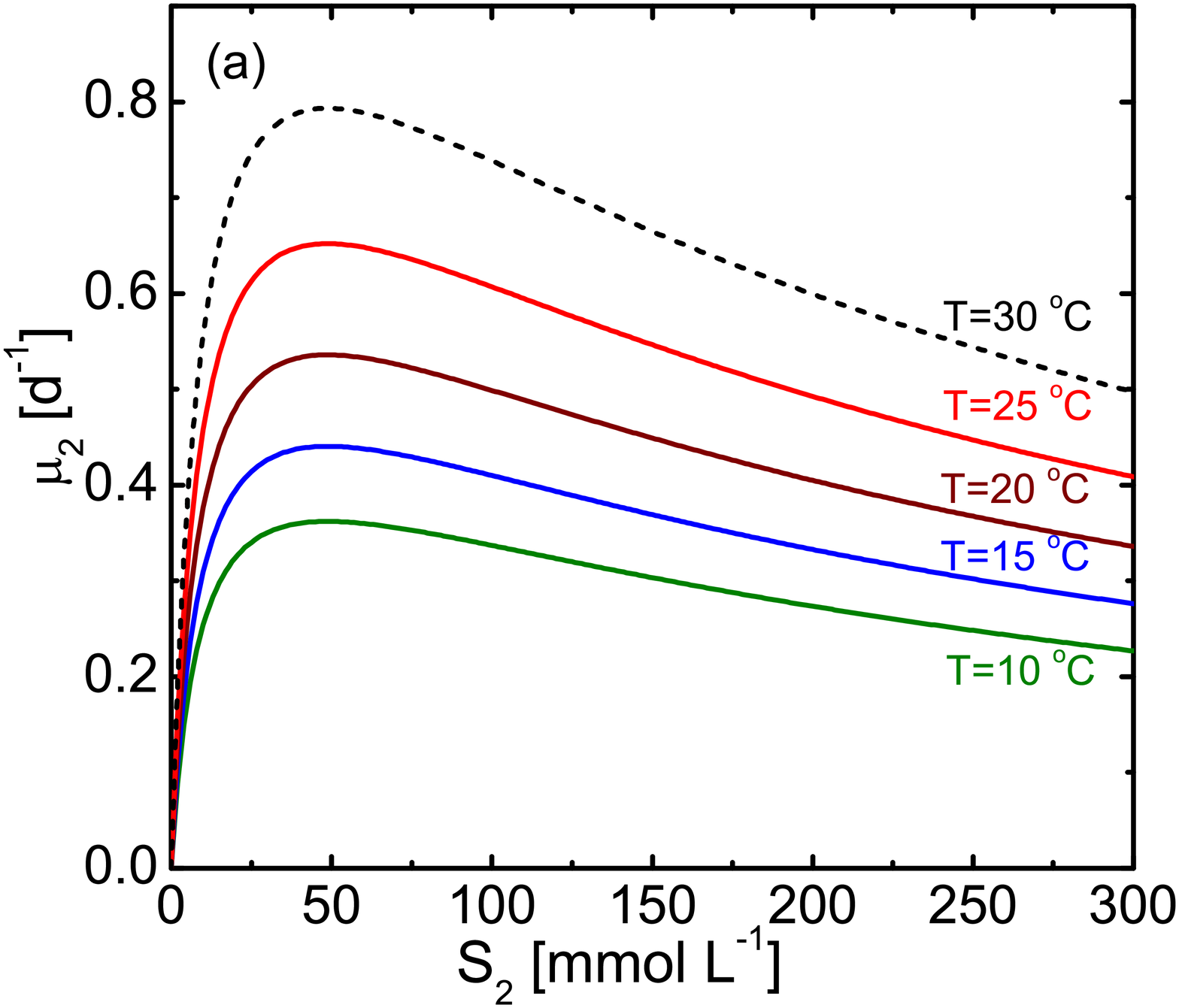}}
  \subfigure{\label{Fig3b}
  \hspace*{-0.3cm}\includegraphics[width=7.8cm,height=7cm]{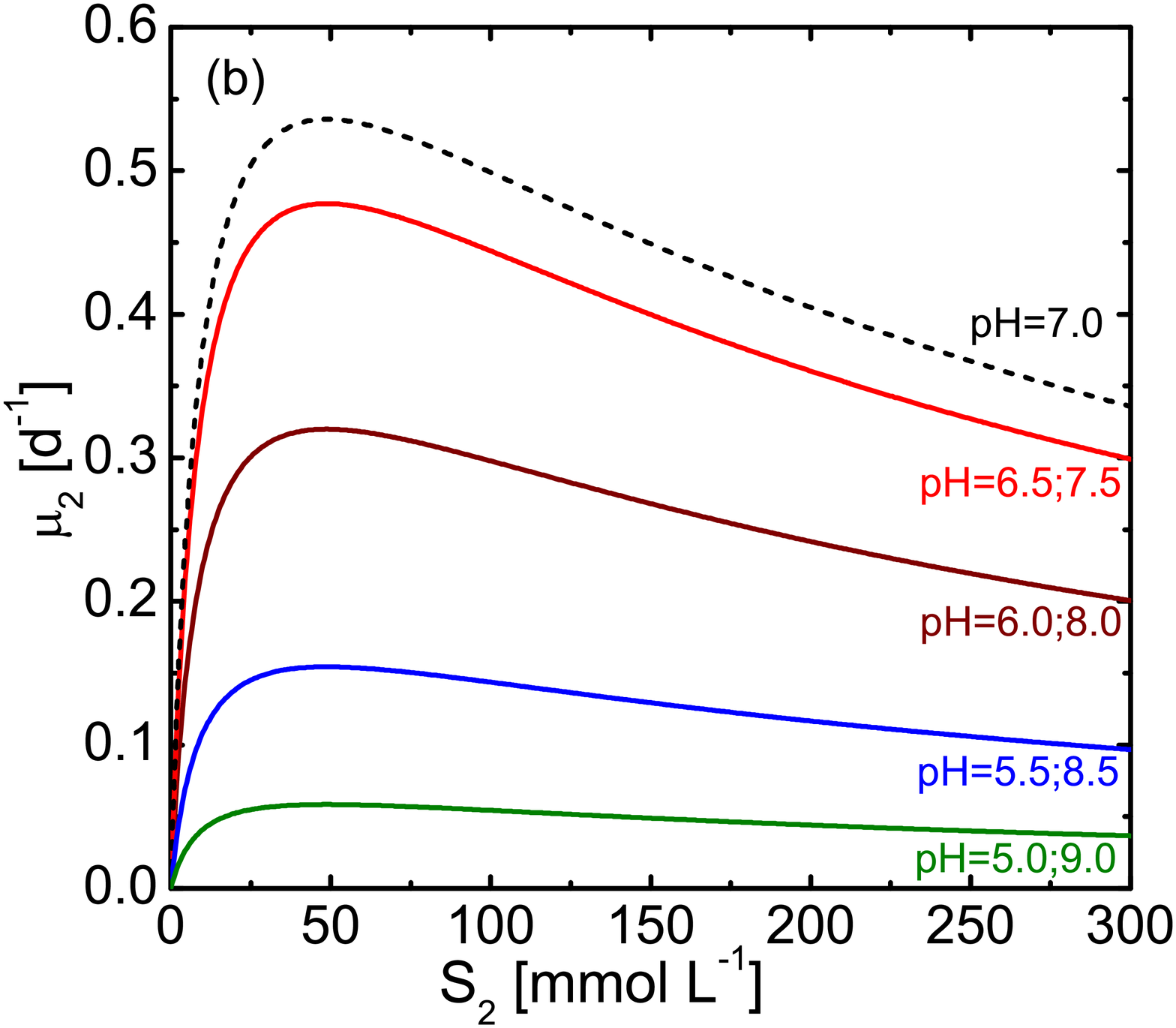}}  
  \caption{
    Haldane growth rate ($\mu_1$) as a function of the organic substrate $S_1$.
  Cases presented:
  (a) changing with the temperature factor $\Theta$,  and 
  (b) changing with the pH factor $I_{\text{pH}}$.  
  The dotted line shows the behaviour of the maximum kinetic growth rates 
  at optimal temperature and pH ({\it i.e.}, at T=30$^\circ$C and pH=7). 
    }
  \label{Fig3}
\end{figure*}

Fig. \ref{Fig3} shows the behaviour of the Haldane growth rate as a function of VFA concentration, temperature and pH. This growth rate model is used in the methanogenic stage and shows the following behaviour: $i)$ increasing function in the range $0\ge S_2 \ge S^{\text{max}}_2 $ with a maximum value at $\bar \mu_2(S^{\text{max}}_2)$ and
$ii)$ decreasing function at $S_2>S^{\text{max}}_2$. Fig. \ref{Fig3a} shows the Haldane growth rate ($\bar \mu_2$), changing with the temperature factor ($\Theta$) at 10, 15, 20, 25, 30$^\circ$C temperatures. Here, a maximum curve of values is observed at 30$^\circ$C corresponding to the optimal temperature. Behind of these temperatures, the tendency of the curves is decreasing. The effect of the pH factor ($I_{pH}$) over the Haldane growth rate model is depicted in Fig. \ref{Fig3b}. Here, the maximum curve of biomass growth rate is obtained at the optimal value of pH ({i.e.}, pH=7.0) and the behaviour of the growth rate is decreasing at values away from the neutrality. 

The simulations were performed using the kinetic parameters values reported by Bernard \textit{et al.} \citep{Bernard01} (see Table \ref{param}).
\subsection{Analytical solutions of equilibrium points}\label{SectionIVa} 
The solutions of the system must be real positive and the feasible values of the state variables $X_1$, $X_2$, $S_1$, $S_2$ satisfy the following qualitative properties:
\begin{equation} \label{cond1}
{X_1} \ge 0, \hspace{0.3cm} {X_2} \ge 0,
\end{equation}
\begin{equation} 
{S_1} \le S_1^0,   \hspace{0.3cm}  {S_2} \le S_2^0 + \frac{{{k_2}}}{{{k_1}}}S_1^0.
\end{equation}
Thus, the following analytical expressions for the equilibrium are obtained, all of them assuming inlet biomass concentration equal zero ($X_1^0=X_2^0=0$).
\subsubsection*{\bf Equilibrium 1: (washout condition)}
The washout condition is given by the absence of biomass in both acidogenic and methanogenic stages, {i.e.}, ${S_1^{*}} = S_1^0$, ${X_1^{*}} = 0$,
 ${S_2^{*}} = S_2^0$, ${X_2^{*}} = 0$. 
\subsubsection*{\bf Equilibrium 2: (washout by acidogenic biomass)} 
The washout condition is given by biomass absence in the acidogenic stage, {i.e.}, ${S_1^{*}} = S_1^0$ and ${X_1^{*}} = 0$. Thus, the equilibrium points associated to the methanogenic stage are 
\begin{equation}
S_2^{*} = \frac{-b \pm \sqrt{b^2-4 a c}}{2 a},
\end{equation}
and
\begin{equation}
X_2^{*} = \frac{ \alpha D (S_2^{*} K_I + S_2^{*} S^2_0 + K_2 K_I) - S_2^{*} \hat{\mu}_2 K_I } {S_2^{*} D \alpha ^2 k_3},
\end{equation}
where $a = \alpha D$, $b=(\alpha D - \hat{\mu}_2) K_I$, $c=\alpha D K_2 K_I$, and $\hat{\mu}_2=\mu_{2\max} \Theta  {I_{pH}}$. Moreover, $S_2$ has a real solution if $a \neq 0$ and the discriminant is positive or zero, $\Delta \geq 0$, \textit{i.e.}, $b^2-4 a c \geq 0$. 
 \subsubsection*{\bf Equilibrium 3: (washout by methanogenic biomass)}
 The washout by absence of biomass in the methanogenic stage is given by
\begin{align}
S_1^{*} & = \frac{\alpha D K_1}{\hat{\mu}_1 - \alpha D },\\
X_1^{*} & = \frac{ \alpha D (S_1^0 + K_1) -\hat{\mu}_1 S_1^0} {k_1 \alpha(\alpha D -\hat{\mu}_1)},\\
S_2^{*} & = \frac{ \alpha D \varUpsilon -( k_2 S_1^0 + S_2^0 k_1) 
\hat{\mu}_1} {k_1 (\alpha D - \hat{\mu}_1)},\\
X_2^{*} & = 0,
\end{align}
where $\varUpsilon=k_2 S_1^0 + k_2 K_1+ S_2^0 k_1$ and $\hat{\mu}_1= \mu_{1\max} \Theta I_{pH}$.
\subsubsection*{\bf Equilibrium 4: (nontrivial solution)} \label{nontrivial}
The analytical nontrivial solution is given by
\begin{align} 
S_1^{*} & = \frac{\alpha D K_1}{\hat{\mu}_1 - \alpha D},\label{nt1}\\
X_1^{*} & = \frac{\alpha D (S_1^0 + K_1) -\hat{\mu}_1 S_1^0} {k_1 \alpha(\alpha D -\hat{\mu}_1)},\\
S_2^{*} & = \frac{-b \pm \sqrt{b ^2-4 a c}}{2 a}, \label{nt2}\\ 
X_2^*   & = {\frac{S_2^*\alpha D ({\alpha}{D}\gamma  - \xi) + \phi}
{S_2^*{\alpha ^2}D{k_1}{k_3}(\alpha D - \hat{\mu}_1)}} ,
\end{align}
where, $\gamma  = {k_2}(S_1^0 + {K_1}) + S_2^0{k_1} + {k_1}{K_I}$,
$\xi  = S_2^0{k_1}\hat{\mu}_1({K_I} + 1) + {k_2}S_1^0\hat{\mu}_1$,
$\varphi  = S_2^*{k_1}{K_I}\hat{\mu}_2$ and
$\psi  = \alpha D{K_2}{K_I}{k_1}$, and 
$\phi=(\hat{\mu}_1 - \alpha D)\varphi  - \left( {\hat{\mu}_1 + \alpha D} \right)\psi$.
 Notice that the equilibrium points of the bioreactor are function of $D$, $T$ and pH parameters.
\subsection{Bifurcation analysis}\label{SectionIVb} 
This section presents the results of bifurcation analysis assuming $D$, $T$ and pH as bifurcation parameters of the system. In all cases, the simulations were performed using the inlet concentrations and parameters values in Table \ref{param}. Notice that all parameters are real positive. \\

\subsubsection{Effect of dilution rate ($D$)}
\begin{figure*}[h]
  \centering
  \includegraphics[width=7.8cm,height=7cm]{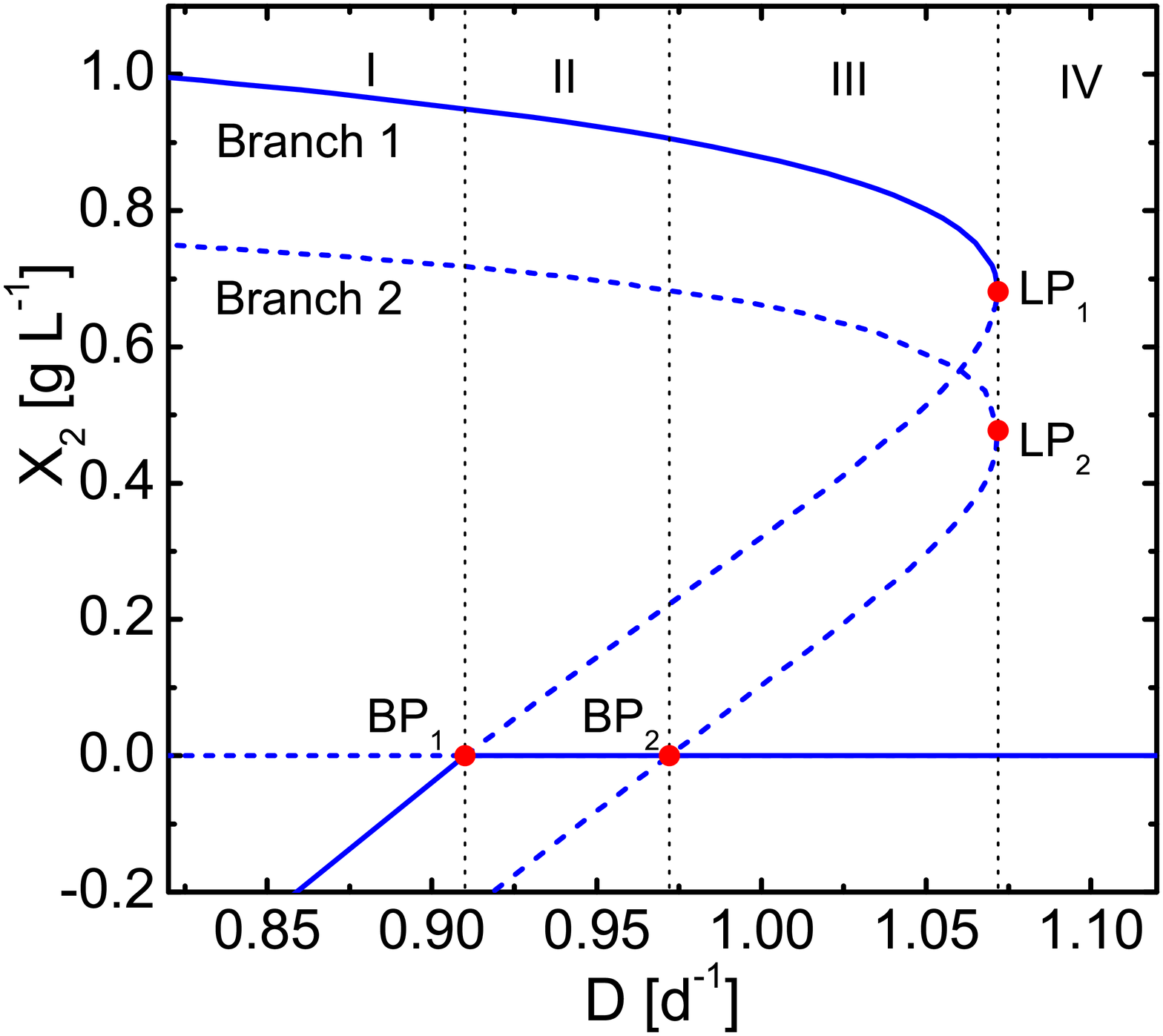}  
  \caption{Bifurcation diagram when changing dilution rate, $D$. 
  Solid lines represent stable equilibrium points and dotted lines represent unstable equilibrium points. 
  BP and LP indicate a branch point bifurcation and limit point, respectively.
  Parameter values used are defined in Table \ref{param}. 
  }
  \label{Fig4}
\end{figure*}

Bifurcation analysis for substrate and biomass concentration has been carried out assuming $D$ as a bifurcation parameter of the system. The continuation in $D$ reaches two limit points (LP) and three branch points (BP). Fig. \ref{Fig4} shows the behaviour of methanogenic biomass concentration $X_2$ changing with $D$. In this figure, solid lines represent stable equilibria and dotted lines represent unstable equilibria. This system exhibits two LP (LP$_1$ and LP$_2$) at $D= 1.0719$. Besides, LP$_2$ is associated to washout of methanogenic bacteria. Similarly, BP$_1$ at $D= 0.9100$ (washout of acidogenic biomass), BP$_2$ at $D= 0.9720$ (washout) and BP$_1$ at $D= 1.6493$ (washout condition). These bifurcation points are qualitatively equivalent to the results reported by Benyahia \textit{et al.} \citep{Benyahia12}. The bifurcation diagrams from the dynamical analysis of the system exhibit a phenomenon denoted as \textit{backward bifurcation}, which involves the existence of a fold bifurcation (LP$_1$) at $D= 1.0719$ and a transcritical bifurcation (BP$_1$) at $D=0.9099$. On the other hand, LP$_2$ shows a second saddle-node bifurcation, which occurs when the critical equilibrium has a null eigenvalue. At the limit point, backward continuation produces the branch point (BP$_2$) at $D=0.9720$. Hence, the bacterial population becomes extinct. Moreover, the bifurcation diagrams show that, depending on the bifurcation parameter value, some equilibrium points may 
exist or not. As can be seen in Fig. \ref{Fig4}, the system is divided and analysed into four regions as a function of $D$ values. In region I ($D<0.9720$), the system shows three equilibrium points with physical meaning, one of them is stable and the others are unstable. In region II ($0.9720<D<1.0719$), the system shows five equilibrium points with physical meaning, two of them are stable and the remaining points are locally asymptotically unstable. In region III, although a stable operational node is presented, it is possible that disturbances affecting the system lead to the other stable equilibrium, which corresponds to the washout condition. Then, in region IV ($D>1.0719$) there are two equilibrium points with physical meaning, one of them is stable and the other one is unstable. Although optimal operating ranges and stability of the system against variations in $D$ have been extensively studied \cite{Hess08,Rincon09,Benyahia12}, a suitable operation range of the bioreactor in Region I at $D<0.91 d^{-1}$ is defined.
\subsubsection{Effect of temperature parameter ($T$)}\label{effecT}
\begin{figure*}[h]
  \centering
\subfigure{\label{Fig5a}
    \hspace*{-1.7cm}\includegraphics[width=7.8cm,height=7cm]{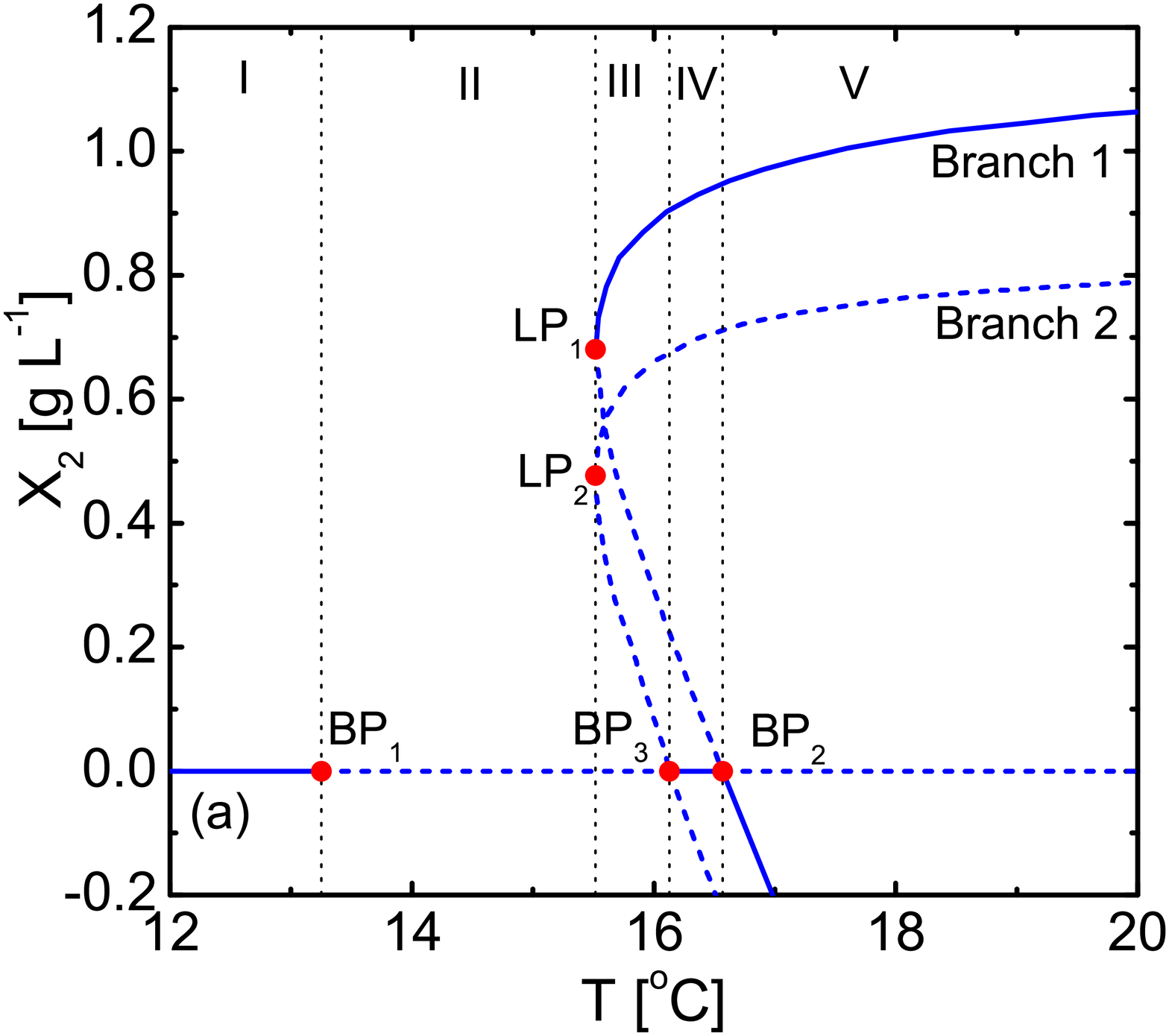}}
   \subfigure{\label{Fig5b}
        \hspace*{-0.3cm}\includegraphics[width=7.8cm,height=7cm]{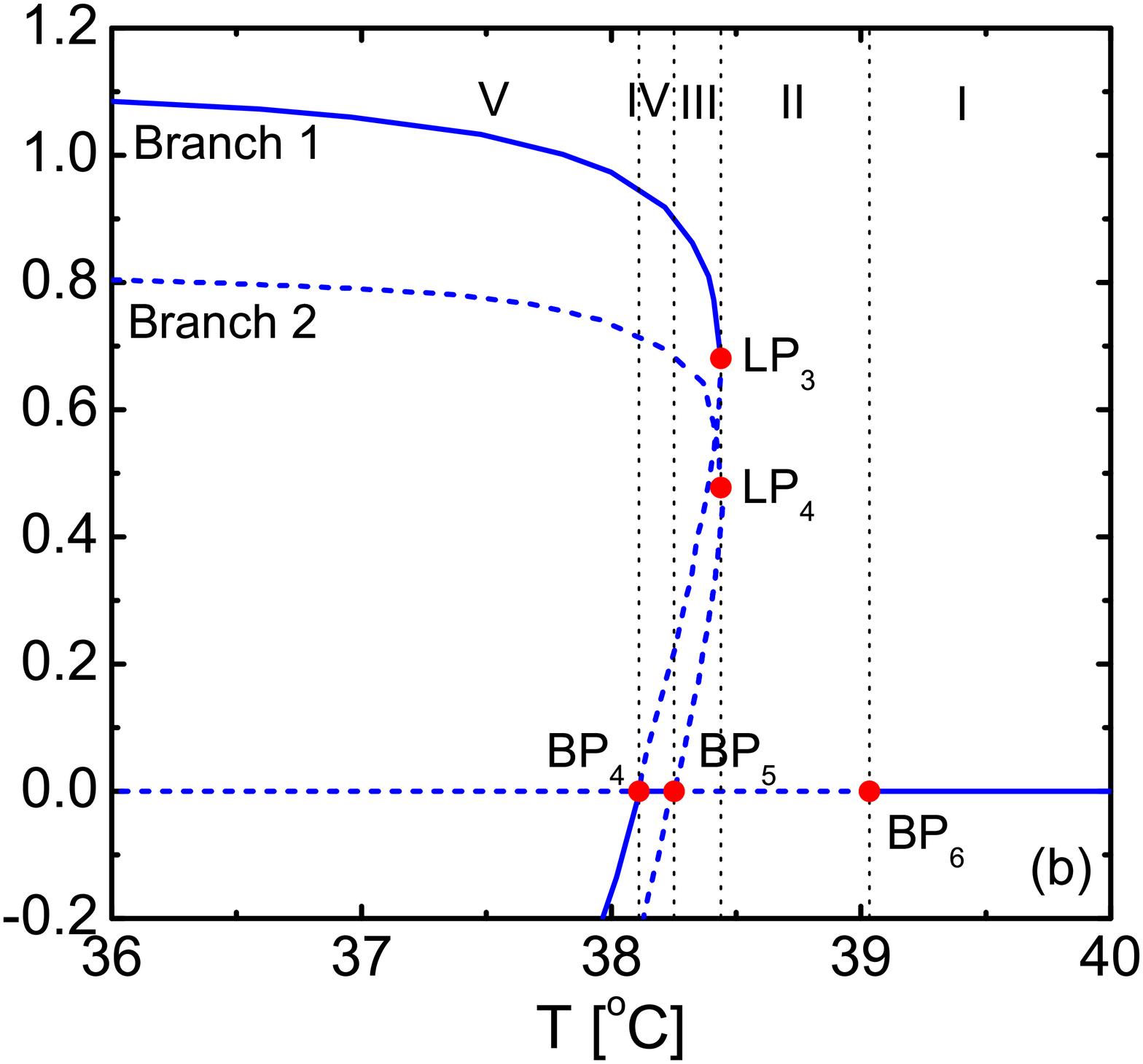} }
  \caption{
  Bifurcation diagram when changing temperature. Solid lines represent state stable equilibrium points and dotted
  lines represent state unstable equilibrium points. BP and LP indicate a branch point bifurcation and limit point,
  respectively. Parameter values used are defined in Table \ref{param}.}
   \label{Fig5}
\end{figure*}

In this section, bifurcation analysis of the system is performed by using $T$ as a bifurcation parameter. In Fig. \ref{Fig5}, methanogenic biomass concentration $X_2$ as a function of the bifurcation parameter $T$ is shown. Here, bifurcation diagram of the system exhibits four LP and six BP with physical meaning. As can be seen in Fig. \ref{Fig5}, the bifurcation diagram exhibits a symmetry in the range of temperatures analysed. In Fig. \ref{Fig5a}, there are two LP denoted by LP$_1$ and LP$_2$ at $T=15.52^{\circ}$C as well as three BP denoted by BP$_1$, BP$_2$ and BP$_3$ at
$T=13.25^{\circ}$C, 16.57$^{\circ}$C and 16.13$^{\circ}$C, respectively. In Fig. \ref{Fig5b}, there are two LP denoted by LP$_3$ and LP$_4$ at $T=38.44^{\circ}$C and also there are three BP denoted by BP$_4$, BP$_5$ and BP$_6$ at $T=38.11^{\circ}$C, 38.25$^{\circ}$C and 39.03$^{\circ}$C, respectively. In Fig. \ref{Fig5}, LP$_2$ and LP$_4$ are associated to washout by acidogenic biomass, and BP$_1$, BP$_3$, BP$_5$ and BP$_6$ points are associated to washout condition.

Fig. \ref{Fig5} shows two-fold bifurcations (LP$_1$ and LP$_3$), where the behaviour of the equilibrium points switches from stable to unstable. At these limit points, a forward continuation leads to the extinction of biomass population of the branch points BP$_2$ and BP$_5$. At these points the system exhibits an exchange stability at $T=$16.57, 38.25$^{\circ}$C by transcritical bifurcation.

The analysis of the number and nature of the equilibria is studied by regions, then, the system is divided into five regions (from I to V) delimited by the bifurcation points obtained. In Region I at $T<$13.25$^{\circ}$C and $T>$39.04$^{\circ}$C, the system shows one equilibrium point with physical meaning, being stable. In Region II (13.25$^{\circ}$C$<T<$15.52$^{\circ}$C and 38.44$^{\circ}$C $<T<$39.04$^{\circ}$C) the system has two equilibrium points with physical meaning, one of them is stable and the other one is unstable. Region III (15.52$^{\circ}$C$<T<$16.13$^{\circ}$C and
38.25$^{\circ}$C$<T<$38.44$^{\circ}$C), shows six equilibrium points with physical meaning, two of them are stable and the others are unstable. In the case of Region IV (16.13$^{\circ}$C$>T>$16.57$^{\circ}$C and 38.11$^{\circ}$C$>T>$38.25 $^{\circ}$C), 
there are five equilibrium points with physical meaning, two of them are locally asymptotically stable and the others are unstable. Finally, in Region V (16.57$^{\circ}$C$>T>$38.11$^{\circ}$C), there are four equilibrium points with physical meaning and only one of them is stable. At this range of temperature, the bacterial growth is favoured, and a stable equilibrium point is guaranteed. These conditions favour a single equilibrium point, which corresponds to normal operating conditions. Therefore, the washout condition by temperature effects is avoided.
\subsubsection{Effect of potential hydrogen parameter (pH)}
\begin{figure*}[h]
  \centering
  \subfigure{\label{Fig6a}
    \hspace*{-1.7cm}\includegraphics[width=7.8cm,height=7cm]{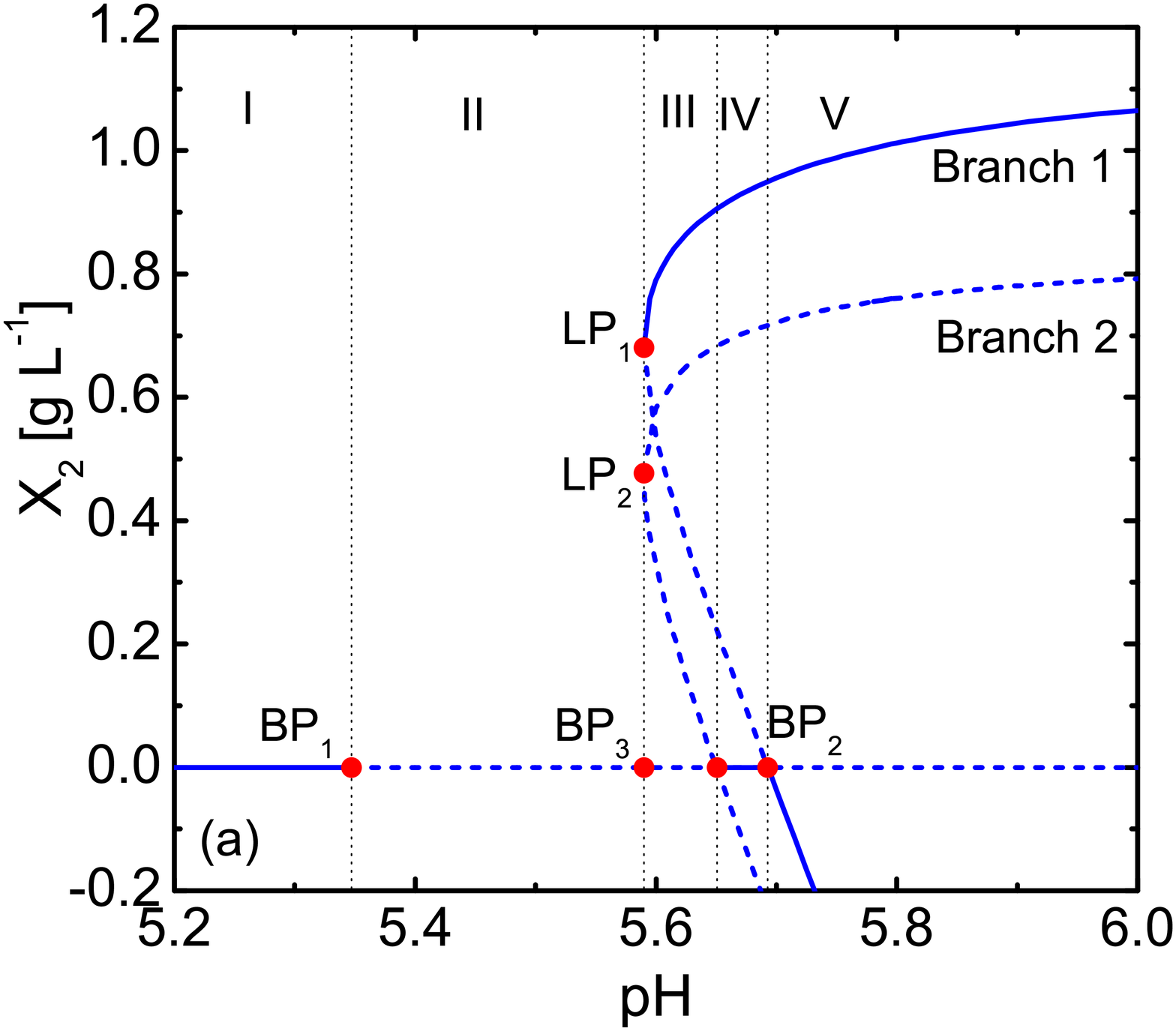}}
   \subfigure{\label{Fig6b}
  \hspace*{-0.3cm}\includegraphics[width=7.8cm,height=7cm]{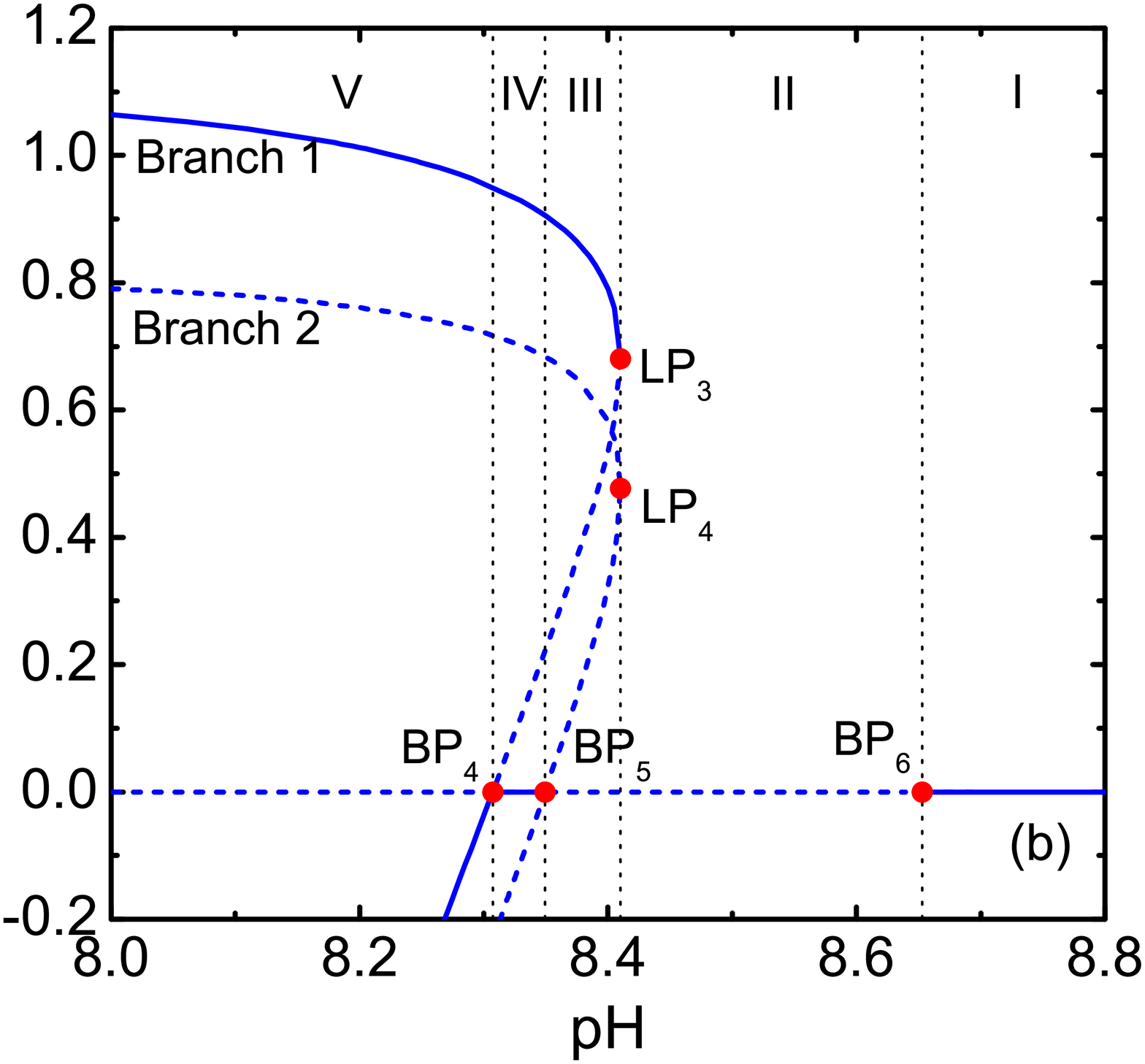}} 
  \caption{Bifurcation diagram when changing pH. Solid lines represent state stable equilibrium points and
  dotted lines represent state unstable equilibrium points. BP and LP indicate a branch point bifurcation and
  limit point, respectively. Parameter values used are defined in Table \ref{param}.}
   \label{Fig6}
\end{figure*}

In this section, pH is used as a bifurcation parameter of the system behaviour. Here, four LP and six BP bifurcation points with physical meaning are observed. The methanogenic biomass concentration $X_2$ changing with the pH parameter is presented in Fig. \ref{Fig6}. Fig. \ref{Fig6a} shows two LP points denoted by LP$_1$ and LP$_2$ at pH=5.59 and also there are three BP denoted by BP$_1$, BP$_2$ and BP$_3$ at pH=5.35, 5.69 and 5.63, respectively. In Fig. \ref{Fig6b} there are two LP denoted by LP$_3$ and LP$_4$ at pH=8.41, as well as three BP denoted by BP$_4$, BP$_5$ and BP$_6$ at pH=8.31, 8.35 and 8.65, respectively. The behavioural analysis of the system leads to identify the existence of two fold bifurcations LP$_1$ and LP$_3$ at pH$=$5.59 and pH$=$8.41, respectively. At these limit points, forward and backward continuation produces the transcritical bifurcations BP$_2$ and BP$_4$ at pH=5.69 and pH= 8.31. 

In Fig. \ref{Fig6}, the system is divided into five regions (from I to V) as follows. 
In Region I (pH$<5.35$ and pH$>8.65$), the system present one equilibrium point with physical meaning and stable. In Region II ($5.35<$pH$<5.59$ and $8.41>$pH$>8.65$), the system shows two equilibrium points with physical meaning, one of them is locally asymptotically stable and the other one unstable. In region III ($5.59<$pH$<5.65$ and $8.35>$pH$>8.41$), there are six equilibrium points with physical meaning, two of them are stable and the others locally asymptotically unstable. In Region IV ($5.65>$pH$>5.69$ and  $8.31>$pH$>8.35$), the system exhibits five equilibrium points with physical meaning, two equilibria are stable and the others unstable. Finally, in Region V ($5.69>$pH$>8.31$), four equilibrium points with physical meaning are determined, only one of them locally asymptotically stable. At this range of pH, the bacterial growth is favoured, and the washout condition is avoided. Then, in Region V suitable operational conditions for anaerobic digestion process are given.\\

The equilibrium points in terms of $\Theta$ and $I_{pH}$ are summarized in Table \ref{equilibrios}. In this table, the stability conditions and feasible equilibrium points are computed according to the established regions in the bifurcation analysis. Notice that, at Regions III and IV, the system exhibits two stable stationary solution competing. The former, corresponds to a nontrivial solution of the system while the latter with a washout by methanogenic biomass. Figures \ref{Fig5} and \ref{Fig6} show how the equilibrium curves \textit{Branch 1} and \textit{Branch 2} cross each other and change their direction at the bifurcation points denoted by LP. Indeed, the \textit{Branch 1} shows the qualitative change of the system properties under quantitative parameter variation at a named fold bifurcation (LP$_1$ and LP$_3$) and at transcritical bifurcations (BP$_2$ and BP$_4$). The transcritical bifurcations are associated to the washout condition. Notice that the values between LP$_1$ and BP$_2$, and LP$_3$ and BP$_4$, are quite close. Therefore, as limit operational values are considered the BP$_2$ and BP$_4$ bifurcation points. At these bifurcation point values the conditions that leads to the system collapse are avoided. Then, from bifurcation analyses a normal operation value of temperature and pH in terms of $\Theta$ and $I_{pH}$ should be between 0.37 and 1.
\subsection{Overloading tolerance of the bioprocess}\label{SectionIVc}
\subsubsection{Acidogenic stage}
\begin{figure*}[h]
  \centering
     \subfigure{\label{Fig7a}
    \hspace*{-1.7cm}\includegraphics[width=7.8cm,height=7cm]{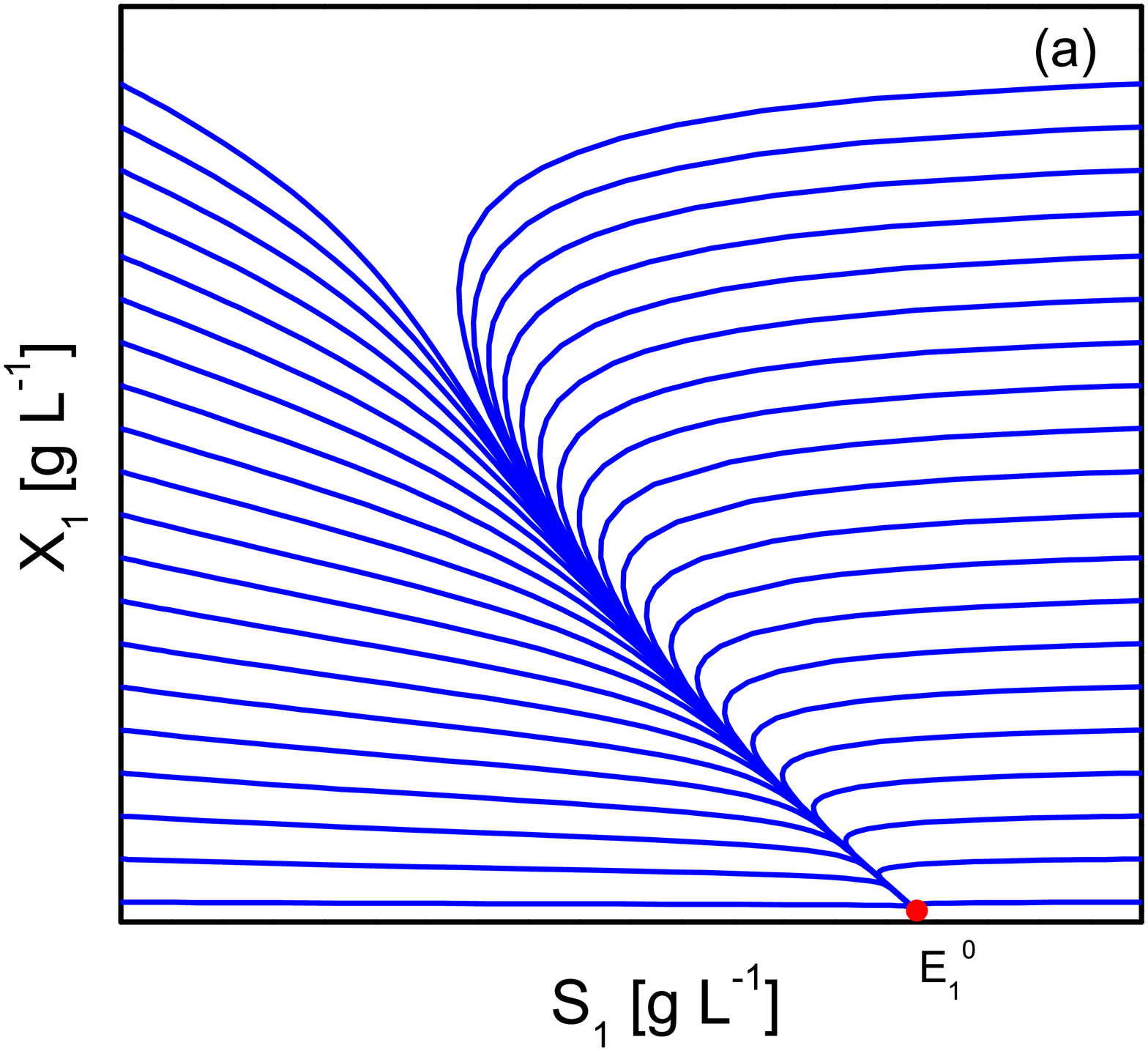} }
     \subfigure{\label{Fig7b} 
    \hspace*{-0.3cm}\includegraphics[width=7.8cm,height=7cm]{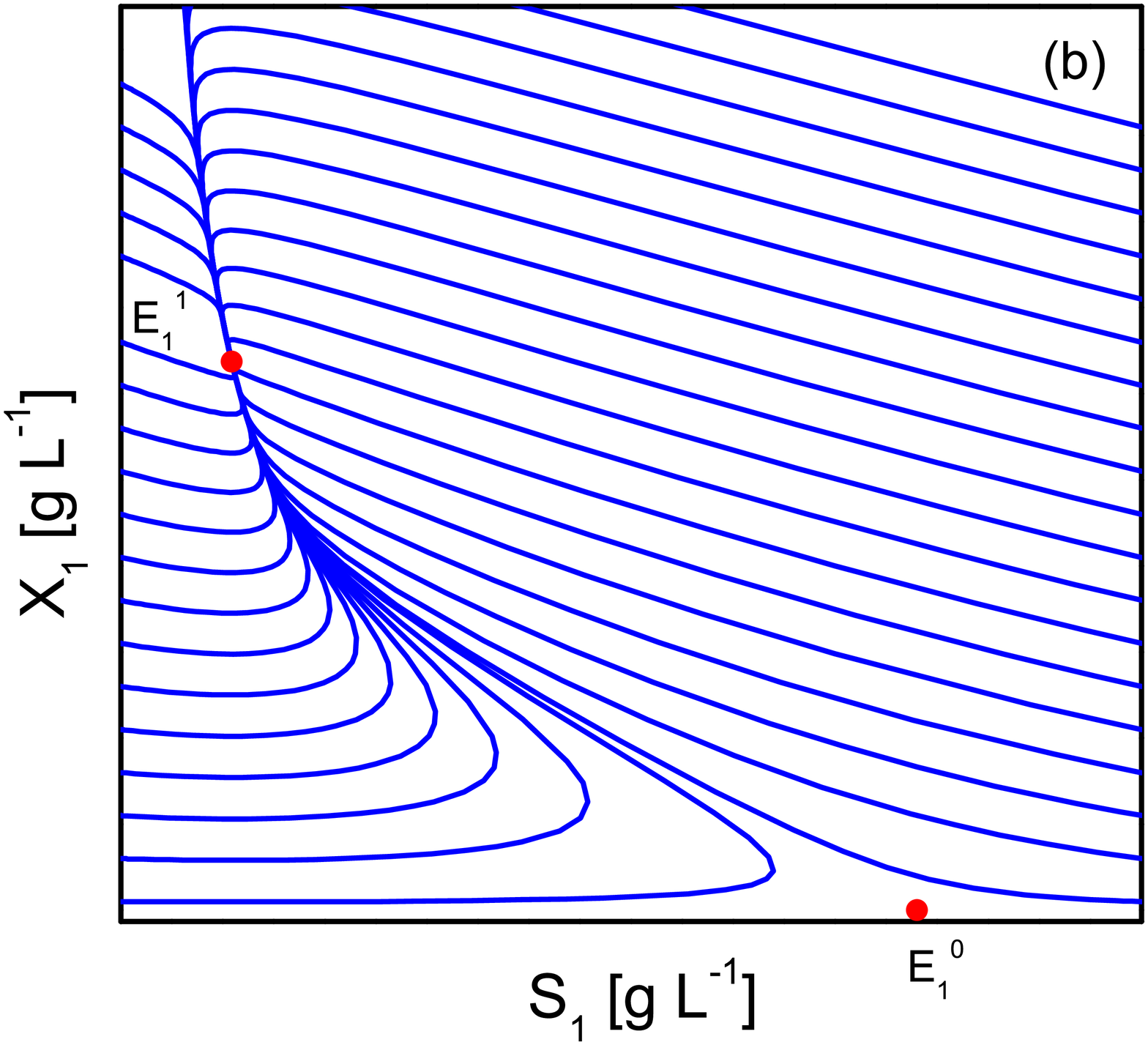} }
  \caption{Phase diagram where E$_1^0$ corresponds to the washout of X$_1$ and E$_1^1$ is the operating stable equilibrium point of acidogenic stage. 
  (a) $\Theta I_{pH}<$0.37 (S$_1^*$=S$_1^0$),
  (b) $\Theta I_{pH}\geq$0.37 (S$_1^*<S_1^0$).
  }
   \label{Fig8}
\end{figure*}

Equations (\eqref{eq3})-(\eqref{eq4}) with $\dot{X_1}=\dot{S_1}=0$ are solved and two cases were studied. In the first case at $\Theta I_{pH}<$0.37, the acidogenic stage exhibits a unique solution with physical meaning, a stable equilibrium point (E$_1^0$), {\it i.e.,} S$_1^*$=S$_1^0$ and X$_1^*=0$. According to Fig. \ref{Fig7a}, this case corresponds to a trivial solution to the washout condition by absence of acidogenic biomass. In the second case, the acidogenic stage exhibits two equilibrium points: one of them is a stable equilibrium point (E$_1^1$) and the other one an unstable equilibrium point (E$_1^0$). Moreover, E$_1^1$ at $\Theta I_{pH}\ge$0.37 corresponding to a nontrivial solution (see Sec. \ref{nontrivial}) of the set equations from the acidogenic stage, {\it i.e.,} S$_1^*<S_1^0$ and X$_1^*>$0  (see Fig. \ref{Fig7b}). Thus, from (\eqref{nt1}), it is defined that the stable operating point E$_1^1$ in terms of the parameters is guaranteed as long as the condition
\begin{equation}\label{condition1}
  \text{Condition 1:} \hspace{0.3cm}  \frac{\hat{\mu}_1}{D} > \alpha . 
\end{equation}
holds.
\subsubsection{Methanogenic stage}
\begin{figure*}[h]
  \centering
    \subfigure{\label{Fig8a}
    \hspace*{-1.7cm}\includegraphics[width=7.8cm,height=7cm]{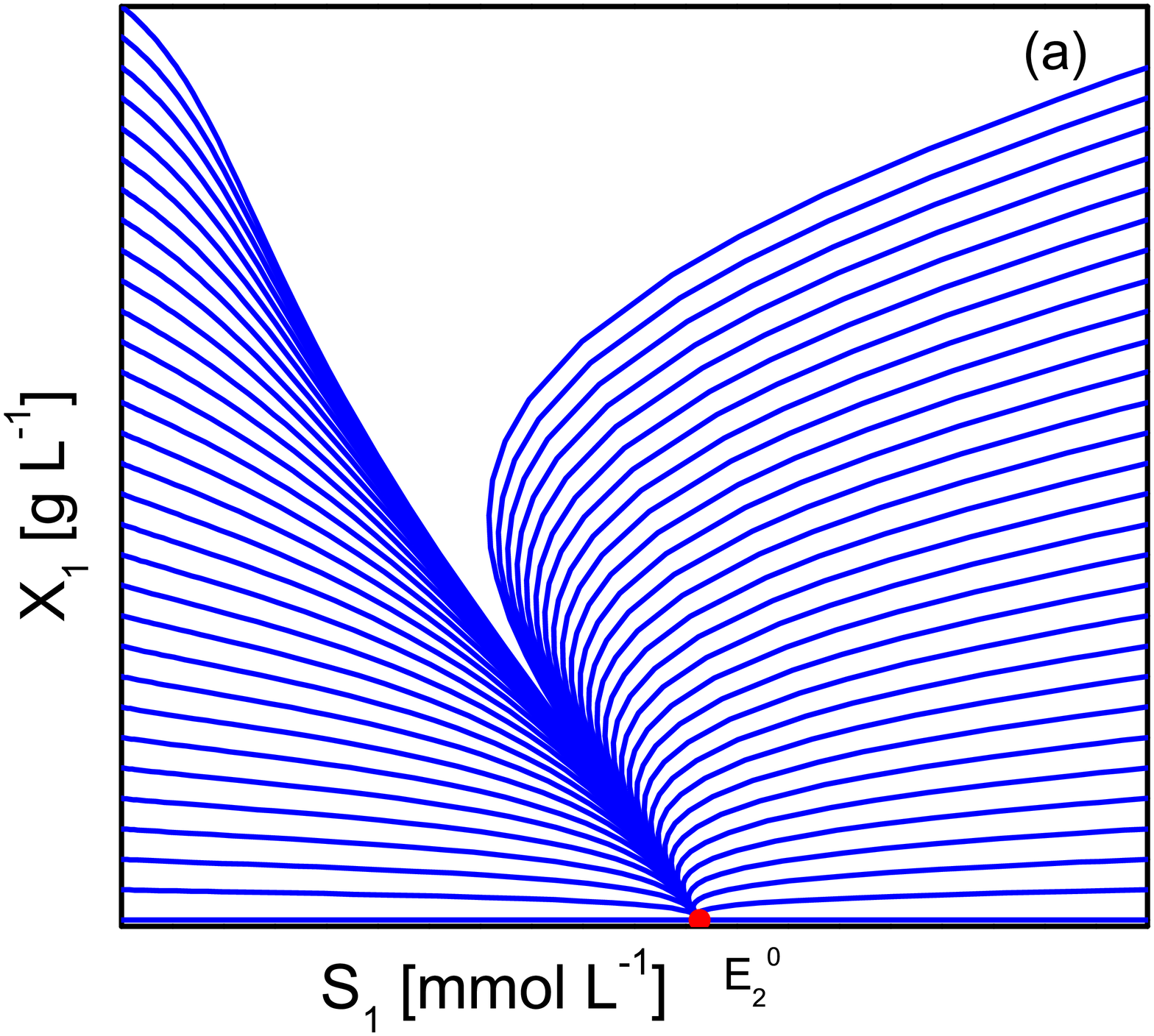}  }
    \subfigure{\label{Fig8b}
    \hspace*{-0.3cm}\includegraphics[width=7.8cm,height=7cm]{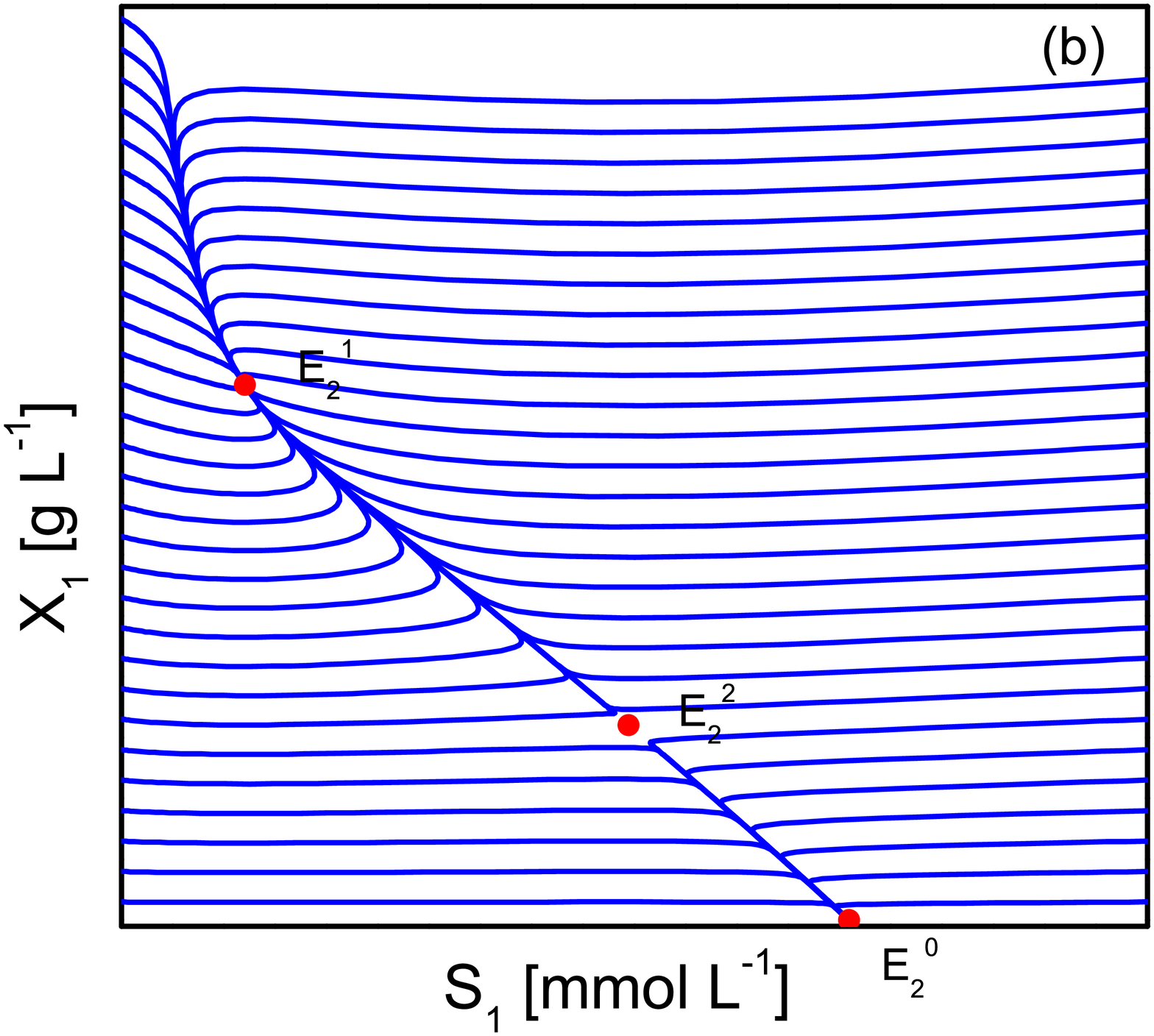}  }\\
    \subfigure{\label{Fig8c} 
    \includegraphics[width=7.8cm,height=7cm]{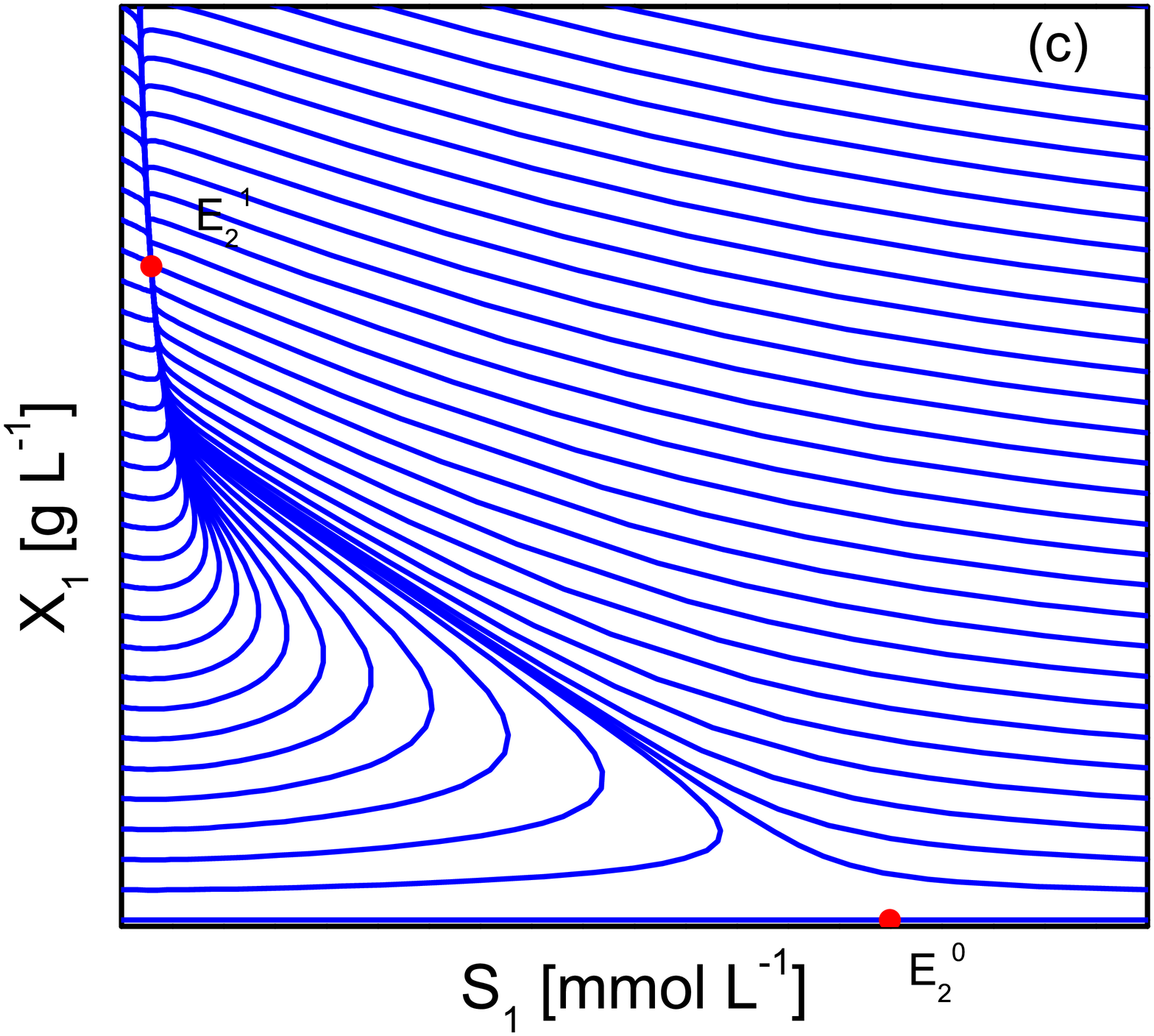}  }
  \caption{Phase diagram where E$_2^0$ corresponds to the washout of X$_2$, E$_2^1$ is the operating stable equilibrium point of methanogenic stage and E$_2^2$ is the unstable equilibrium point of acidogenic stage. 
  (a) $\Theta I_{pH}<$0.37 (S$_2^*$=S$_2^0$),
  (b) 0.37$\leq \Theta I_{pH} <$0.64 (S$_2^* \le S_2^0 + \frac{k_2}{k_1} S_1^0$),
  (c) $\Theta I_{pH} \geq$ 0.64 (S$_2^* < S_2^0 + \frac{k_2}{k_1} S_1^0$).}
   \label{Fig9}
\end{figure*}

In this section the methanogenic stage is studied. The simulation was carried out for S$_1^*<S_1^0$ and X$_1^*>$0, which correspond to an stable operational point different to the washout condition. The curves in Fig. \ref{Fig9} show the phase-diagram results for the system of differential equations given by (\eqref{eq5})-(\eqref{eq6}) with $\dot{X_2}=\dot{S_2}=0$. From the solution of these differential equations, three particular solution cases are obtained: the first case $\Theta I_{pH} <$0.37, the methanogenic stage shows a stable equilibrium point (E$_2^0$), when S$_2^*$=S$_2^0$ and X$_2^*=$0. This equilibrium point corresponds to the washout condition by absence of methanogenic biomass. The phase diagram is depicted in Fig. \ref{Fig8a}. In the second case 0.37$\leq \Theta I_{pH} <$0.64, the methanogenic stage shows three equilibrium points with physical meaning (E$_2^0$, E$_2^1$ and E$_2^2$). The equilibrium points E$_2^1$ and E$_2^2$ corresponds to a nontrivial solution at S$_1^*<S_1^0$ and S$_2^* \le S_2^0 + \frac{k_2}{k_1} S_1^0$. As can be observed in Fig. \ref{Fig8b}, the equilibrium points E$_2^0$ and E$_2^1$ are asymptotically stable.  
Finally, in the third case $\Theta I_{pH} \geq 0.64$, the methanogenic stage shows two equilibrium points (E$_2^0$ and E$_2^1$). The results of this case are shown in Fig. \ref{Fig8c}. The stable equilibrium point E$_2^1$ corresponds with a nontrivial solution, which guarantees a stable and optimal operation condition of the process.
Then, the stability and operability of the system are guaranteed when the following system condition is taken into account:
\begin{equation}\label{condition2}
  \text{Condition 2:} \hspace{0.3cm} \left( \frac{ \alpha D - 
  \hat{\mu}_2}{D} \right)^2  \ge {\frac{ 4 K_2 \alpha ^2}{K_I}}.
\end{equation}
\subsection{General operational indicator control-oriented risk index} \label{SectionIVd}  
Based on the previous dynamical analyses, it is shown that the stability of the system depends on certain conditions that could be measurable in terms of the values of $D$, $T$ and pH. In fact, from previous sections, it was determined that, while conditions defined in Sec. \ref{SectionIVb} guarantee the operability of the system, some other conditions guarantee the stability of the process. 

Therefore, an effective on-line monitoring and control parameter during the anaerobic digestion process is the accomplishment of Conditions 1 and 2 given by (\eqref{condition1}) and (\eqref{condition2}), respectively. Then, a risk index is proposed in order to warn in a preventative way a system destabilization and hence, the washout condition. A monitoring strategy that minimizes the risk of reactor
destabilization due to $T$, pH and $D$ effects is given by 
\begin{equation}\label{SRI}
 SRI=
\left\{
\begin{aligned}
&0~\text{for}~\frac{\hat{\mu}_1}{D} >  \alpha ~\&~\left( \frac{ \alpha D - 
  \hat{\mu}_2}{D} \right)^2  \ge {\frac{ 4 K_2 \alpha ^2}{K_I}},\\
&1~\text{for}~\frac{\hat{\mu}_1}{D}\le \alpha ~\&~
\left( \frac{ \alpha D -  \hat{\mu}_2}{D} \right)^2  < {\frac{ 4 K_2 \alpha ^2}{K_I}},\\
\end{aligned}
\right.
\end{equation}
where both $\hat \mu_1$ and $\hat \mu_2$ are function of $T$ and pH. 

If Conditions 1 and 2 are satisfied, there is no of risk (SRI$=0$), therefore, the reactor is in a normal and stable operational area. Otherwise, if those conditions are not satisfied, the risk-index \eqref{SRI} indicates that the behaviour of the reactor is in undesirable operational area with washout risk (SRI$=1$), and it is necessary to make operational decisions.
\subsubsection{Numerical simulation using a risk index}
\begin{figure*}[h]
    \centering
    \subfigure{\label{Fig9a}
    \hspace*{-1.7cm}\includegraphics[width=7.8cm,height=7cm]{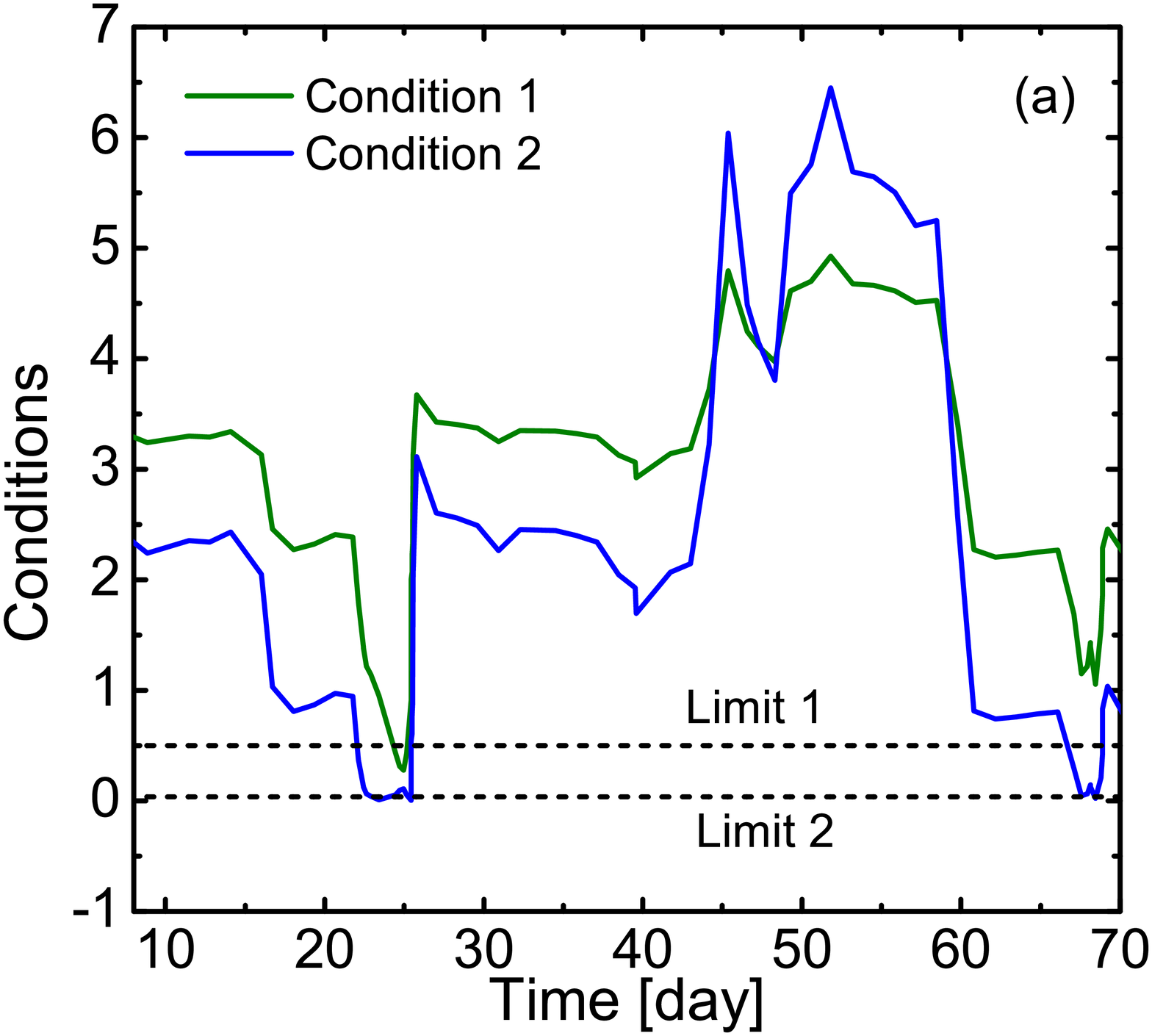}}
    \subfigure{\label{Fig9b}
    \hspace*{-0.3cm}\includegraphics[width=7.8cm,height=7cm]{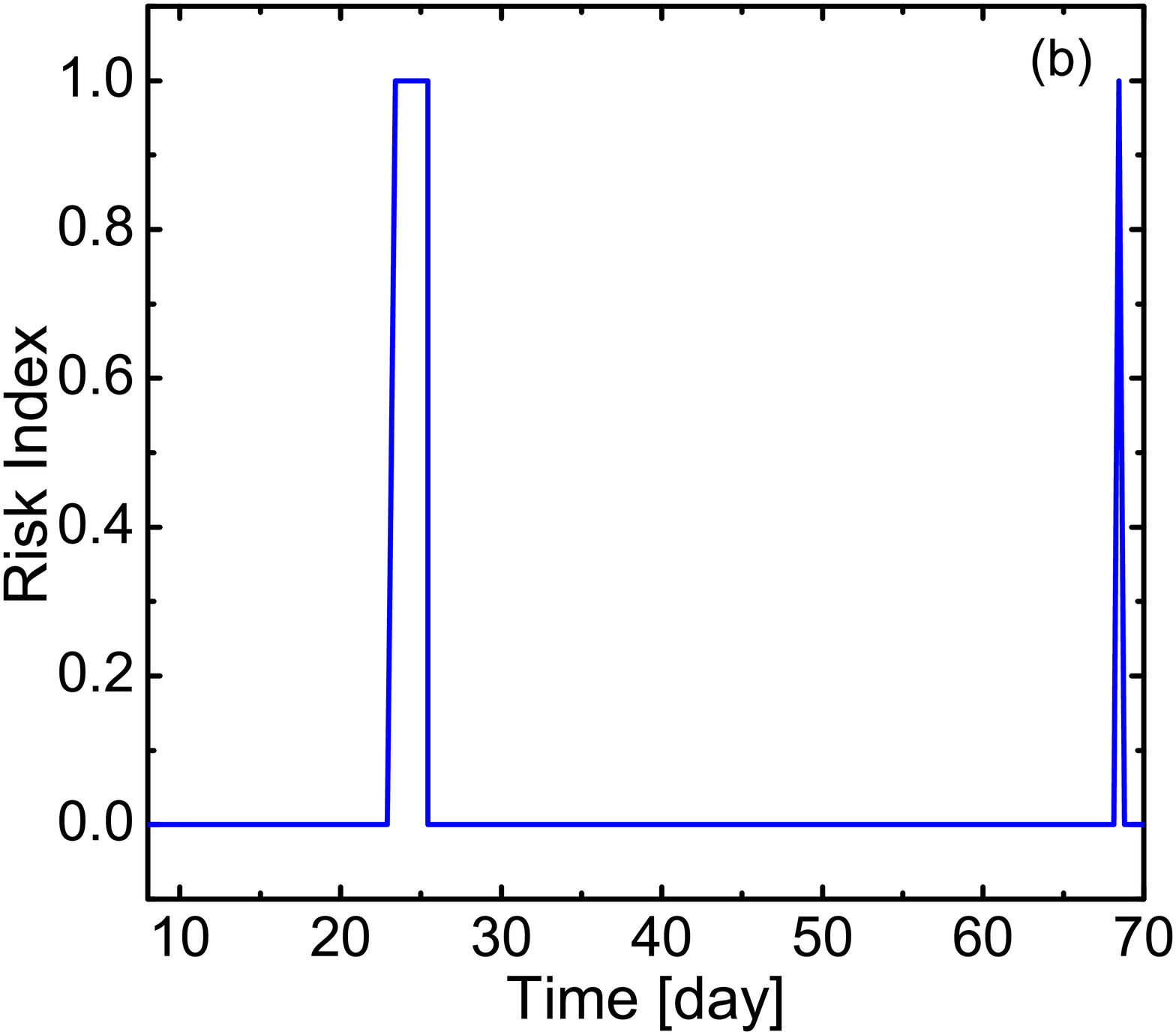}}
  \caption{Operational indicator control-oriented risk index. Cases presented: 
  (a) numerical simulations of Condition 1 in \eqref{condition1} and Condition 2 in \eqref{condition2} 
  using experimental data reported by Bernard \textit{et al.} \cite{Bernard01} and 
  (b) risk-index computations.}
   \label{Figrisk}
\end{figure*}

The risk index is given by (\eqref{SRI}) while the simulations were computed by using $D$ and pH data from Bernard \textit{et al.} \cite{Bernard01}. The numerical results are shown in Fig. \ref{Figrisk}. In this figure, both Conditions 1 and 2 are included. Dotted line is the limit condition that corresponds to the constant value calculated from the right side of the inequalities at \eqref{condition1} and \eqref{condition2}. Solid line corresponds to left side of \eqref{condition1}-\eqref{condition2}. For the sake of simplicity, normalized values are used in Fig. \ref{Fig9a}.

The risk index computations are shown in Fig. \ref{Fig9b}. As can be seen in this figure, 0 is the value used to define stable and normal operational, and 1 is used for
unstable condition. Here, the results are shown as a function of time in order to compare with the behaviour of experimental data reported by Bernard \textit{et al.} \cite{Bernard01}. The non-compliance with the stability conditions means that the system is conducted to destabilization, and possibly led to the washout. As one can notice, the on-line measurement of state variables ($S_1$, $X_1$, $S_2$ and $X_2$)
is difficult, but in this case the risk index only depends on kinetics and the easiness easily of measuring on-line the parameters of the process.

A comparative analysis between the behaviour of the substrate and VFA concentrations reported by Bernard \textit{et al.} \cite{Bernard01}, and the estimation of the risk index (see Fig. \ref{Fig9b}) allows to see alerts when there is an increase in the VFA concentration. These increments are one of the most common causes of destabilization in bioreactors. As it has mentioned before by other authors \cite{Hess08}, high values of the risk index (close to one) are representative of regimes of acid accumulation. 

In this case, a possible acidification process from the dilution rate and pH measurements were detected through the values of the risk index. Then, this index allows to prevent the destabilization of the reactor without direct measurement of the state variables and changes in the measured parameters. The proposed risk index could be applied in real processes using the adequate on-line instrumentation to measure the required parameters. Therefore, the measurements allowing to make a sensibility analysis of the risk index and evaluate its applicability on-line.
\section{Conclusions}\label{SectionV} 
A mathematical model for an anaerobic digester reactor with two main reaction stages
-methanogenic and acidogenic- was presented in this paper. The model takes into account the influence of some important variables such as temperature and pH. A dynamic analysis of non-trivial and biologically feasible steady state using bifurcation theory was performed. The systematic dynamical analysis was developed taking into account variations of temperature and pH on the growth rate models. Regions with multiple steady states, unstable operational points, washout conditions and transcritical bifurcations were found.

Furthermore, a detailed study of the bifurcation branches was performed. The qualitative and quantitative study of the equilibrium points allowed to identify normal and optimal operational conditions in which the operational condition was becoming stable and the washout is avoided. 

In this way, a risk management criterion was established as a function of dilution rate, pH, and temperature measurements. The criterion is based on two conditions established from the dynamic analysis of the system. The purpose of the criterion is to generate an alert when the process evolves into an unstable operational area and at washout risk. This criterion could be used to monitor the operational conditions of the bioprocess on-line and in real time, \textit{i.e.}, the bioreactor could be monitored by on-line sensing devices in situ. In fact, the criterion proposed here is a useful tool for designing a real-time simulator for supervisory control of the process. The implementation in a real process will be require using a commercial instrumentation available and a visualization programming tool. The monitoring system requires the use of more economical instrumentation compared with the actual strategies used nowadays for the state variables measurements. 

The results obtained confirm the importance of dynamical analysis oriented to monitoring, optimization and control of anaerobic digestion processes. The risk-index based criterion proposed in this paper could be used to decide and take the best actions and strategies in order to anticipate future problems in the bioreactor. 

Future research topics are oriented to study more complex systems with multi-stage units, recycle and multiplicity of steady states. The ranges of temperature and pH established here through the bifurcation analysis are the goal of the control design strategies. Work in this direction is currently in progress.
\section*{Acknowledgments}
A.M. Alzate wants to acknowledge the program \textquotedblleft Programa de becas para estudiantes sobresalientes de posgrado\textquotedblright from Universidad Nacional de Colombia - Sede Manizales and the Institut de Rob\'otica i Inform\'atica Industrial (IRI) from the Universitat Polit\'ecnica de Catalunya.
\begin{table*}[h]
\centering
\caption{Inlet concentrations and nominal kinetic parameter values} \label{param}
\begin{tabular}{ccl}
\hline 
Parameter &  Values of Ref. \cite{Bernard01} &  Unit \\
\hline
$k_1$ 		& 42.14 & g COD$/$g X$_1$ \\
$k_2$ 		& 116.5 & mmol VFA$/$g X$_1$ \\
$k_3$ 		& 268 & mmol VFA$/$g X$_2$ \\
$K_{1}$ 	& 7.1  & g$/$L  \\
$K_{2}$ 	& 9.28  & mmol$/$L \\
$K_{I}$ 	& 256  & (mmol$/$L)$^2$ \\
$S_1^0$     & 15.6 & g$/$L \\
$S_2^0$     & 112.7 & mmol$/$L \\ 
$X_1^0$     & 0.0   & g$/$L \\
$X_2^0$     & 0.0   & g$/$L \\
$\alpha$ 	&  0.5 & -- \\
${\mu _{1\max }}$ &  1.2 & d$^{-1}$\\
${\mu _{2\max }}$ &  0.74 & d$^{-1}$\\
[1ex]
 \hline 
\end{tabular} 
\end{table*}
\begin{table*}[h]
\centering
\caption{Stability of equilibrium points of the system} \label{equilibrios}
\begin{tabular}{cclll}
\hline
    Region & Condition & Equilibria &  Lyapunov Stability & Hurwitz Stability Criterion   \\
\hline
I & {$\Theta<0.22$ or} & {$(0,S_1^0,0,S_2^0)$} &  {Stable} & {Ensured} \\
  & {$I_{pH}<0.22$}   &                       &           &          \\
\hline
II &{$0.22<\Theta<0.34$ or} & {$(0,S_1^0,0,S_2^0)$} & Unstable & {Not ensured} \\
   & {$0.22<I_{pH}<0.34$} & {$(X_1^*,S_1^*,0,\hat S_2)$} &  Stable & {Ensured} \\
\hline
III &{$0.34< \Theta< 0.37$ or} & {$(0,S_1^0,0,S_2^0)$} &   {Unstable} & {Not ensured} \\
    & {$0.34<I_{pH}<0.37$} & {$(X_1^*,S_1^*,0,\hat S_2)$} &   {Stable}  & {Not ensured} \\
    &  & {$(X_1^*,S_1^*,X_2^*,S_2^*)$} &   {Stable} & {Ensured} \\
    &  & {$(X_1^*,S_1^*,X_2^*,S_2^*)$} &   {Unstable} & {Not ensured} \\
    &  & {$(0,S_1^0,X_2^*,S_2^*)$} & {Unstable} & {Not ensured} \\
    &  & {$(0,S_1^0,X_2^*,S_2^*)$} &  {Unstable} & {Not ensured} \\
\hline
{IV} & {$0.37<\Theta<0.40$ or} & {$(0,S_1^0,0,S_2^0)$} &  {Unstable} & {Not ensured} \\
     & {$0.37<I_{pH}<0.40$} &{$(X_1^*,S_1^*,0,\hat S_2)$} & {Stable} & {Ensured} \\
     &  & {$(X_1^*,S_1^*,X_2^*,S_2^*)$} &   {Stable} & {Ensured} \\
     &  & {$(X_1^*,S_1^*,X_2^*,S_2^*)$} &   {Unstable} & {Not ensured} \\
     &  & {$(0,S_1^0,X_2^*,S_2^*)$} &  {Unstable} & {Not ensured} \\
\hline
{V} & {$0.40<\Theta \le 1.00$ or} & {$(0,S_1^0,0,S_2^0)$} &   {Unstable}  & {Not ensured} \\
    & {$0.40<I_{pH} \le 1.00$} & {$(X_1^*,S_1^*,0,\hat S_2)$} &   {Unstable} & {Not ensured} \\
    &  &{$(X_1^*,S_1^*,X_2^*,S_2^*)$} &  {Stable} & {Ensured} \\
    &  & {$(0,S_1^0,X_2^*,S_2^*)$} & {Unstable} & {Not ensured} \\
\hline
\end{tabular}  
\end{table*}

\newpage
\clearpage

\appendix
\section{Stability analysis of the system} \label{Stability}
\begin{theorem} \textbf{Lyapunov's Indirect Method}.
Consider a dynamical system defined by 
\[ \dot{x}=f(x), \hspace{0.2cm} x \in \mathbb{R} ^n, \]
where $x$ are the state variables and $f:\mathbb{R}^n \longmapsto \mathbb{R}^n$ is smooth function. Suppose that it has an equilibrium $x^*$, and $A$ denotes the Jacobian matrix of $f(x)$ evaluated at $x^*$. Then, $x^*$ is asymptotically stable if all eigenvalues $\lambda_1$, $\lambda_2$,$\cdots$, $\lambda_n$ of $A$ satisfy $Re({\lambda_i}) < 0$.
\end{theorem}
Thus, the linearisation around $x^*$ leads to the following Jacobian matrix 
\begin{equation} \label{Jacobian}
A = \left[ {\begin{array}{*{20}{c}}
A_{11}&0&A_{13}&0\\
0&A_{22}&0&A_{24}\\
A_{31}&0&A_{33}&0\\
A_{41}&A_{42}&A_{43}&A_{44}
\end{array}} \right],
\end{equation}
where
$A_{11}={\mu _1^*} - \alpha D$, $A_{13}=\mu _1^,{X_1^*}$,
$A_{22}={\mu _2^*} - \alpha D$, $A_{24}=\mu _2^,{X_2^*}$,
$A_{31}={ - {k_1}{\mu _1^*}}$, $A_{33}= -{k_1}\mu _1^,{X_1^*} - D$,
$A_{41}={{k_2}{\mu _1^*}}$, $A_{42}={-{k_3}{\mu _2}}$, $A_{43}={{k_2}\mu _1^,{X_1^*}}$, 
and  $A_{44}=-{k_3}\mu _2^,{X_2^*} - D$. Moreover, ${\mu _1^*}$, ${\mu _2^*}$ are growth rate kinetic model evaluated at $S_1^*$ and $S_2^*$, respectively.
Likewise, $\mu _1'$ is the derivative of $\mu_1$ with respect to $S_1^*$ and $\mu _2'$ is the derivative of $\mu_2$ with respect to $S_2^*$ both computed as follow:
\begin{equation}  \label{dniu1}
\mu _1' = \frac{{{\mu _{1\max }}\Theta {I_{pH}}{K_1}}}{{{{\left( {{K_1} + {S_1^*}} \right)}^2}}},
\end{equation}
\begin{equation} \label{dniu2}
\mu _2' = \frac{{{\mu _{2\max }}\Theta {I_{pH}}{K_I}\left( {{K_2}{K_I} - {S_2^*}^2} \right)}}
{{{{\left( {{K_2}{K_I} + {S_2^*}{K_I} + {S_2^*}^2} \right)}^2}}}.
\end{equation}
After perform some algebraic operations, the Jacobian eigenvalues are given by
\begin{equation}
{\lambda _{1,2}} = {
\frac{1}{2}({A_{44}} + {A_{22}}) \pm \frac{1}{2}\sqrt {{{({A_{44}} - {A_{22}})}^2} + 4{A_{42}}{A_{24}}}},
\end{equation}
\begin{equation}
{\lambda _{3,4}} = {
\frac{1}{2}({A_{33}} + {A_{11}}) \pm \frac{1}{2}\sqrt {{{({A_{33}} - {A_{11}})}^2} + 4{A_{31}}{A_{13}}}}.
\end{equation}
The stability analysis around the equilibrium point using the Lyapunov's indirect 
method is restricted to infinitesimal neighbourhoods of the equilibrium point \cite{Slotine91},\textit{ i.e.}, this method is a necessary condition but not sufficient in order to ensure the stability of the system. Hence, the Hurwitz criterion \cite{Allen07} is used to establish some conditions with the range of values from the parameters ensuring system stability conditions. 
\begin{theorem} \textbf{Routh-Hurwitz Criterion}. 
Given the polinomial 
$P(\lambda)=\lambda^n+a_1\lambda^{n-1}+\cdots+a_{n-1}\lambda+a_n$, where the coefficients $a_i$ are real constants, $i=1,\cdots,n$, define the $n$ Hurwitz matrices using the coefficients $a_i$ of the characteristic polynomial. All of the roots of the characteristic polynomial have negative real part if and only if the determinants of all Hurwitz matrices are positive.
\end{theorem}
Thus, the characteristic equation of the Jacobian matrix related to the system is given by \[ \lambda^4+a_1\lambda^3+a_2\lambda^2+a_3\lambda+a_4=0,\] 
where the coefficients are $a_1= -A_{44}-A_{33}-A_{22}-A_{11}$,
$a_2= -A_{42} A_{24}+A_{44}(A_{33}+A_{22}+A_{11})-A_{31}A_{13}+A_{33}(A_{22}+A_{11})+A_{22}A_{11}$, $a_3= A_{42}A_{24}(A_{33}+A_{11})+A_{44}(A_{31}A_{13}-A_{33}A_{22}-A_{33}A_{11}-A_{22}A_{11})+A_{22}(A_{31}A_{13}-A_{33}A_{11})$, and 
$a_4=A_{42}A_{24}(A_{31}A_{13}-A_{33}A_{11})-A_{44}A_{22}(A_{31}A_{13}+A_{33}A_{11}).$
In order to ensure that $Re({\lambda _i}) < 0$, for all $i$, it is considered each term independently and it is verified, according to the Routh-Hurwitz simplified criterion, that $a_1>0$, $a_3>0$, $a_4>0$, and $a_1a_2a_3>a_3^2+a_1^2a_4$ \cite{Allen07}. \\
\section*{References}
\bibliography{biblio}
\bibliographystyle{rsc}
\end{document}